\documentclass{aa}
\usepackage{graphicx}
\usepackage{amsmath}
\usepackage[varg]{txfonts}
\usepackage{makecell}
\usepackage{booktabs}
\usepackage{hyperref}
\usepackage{natbib}
\usepackage{multirow}
\usepackage{xcolor}

\begin{document}

\title{Evidence for a Fast, Long-Period, Hot Population of Accreting Neutron Stars}

\author{
A.~Patruno\inst{1,2}
\and
M.~Ferrer-Martinez\inst{3,4}
\and
B.~Haskell\inst{4,5}
\and
Y.~Cavecchi\inst{6,7}
}

\institute{
Institute of Space Sciences (ICE, CSIC), Campus UAB, Carrer de Can Magrans s/n, E-08193 Barcelona, Spain
\and
Institut d’Estudis Espacials de Catalunya (IEEC), E-08034 Barcelona, Spain
\and
Departament de Física, Universitat de les Illes Balears, IAC3 – IEEC, Crta. Valldemossa km 7.5, E-07122 Palma, Spain
\and
Università di Milano, Dipartimento di Fisica, Via Celoria 16, 20133 Milano, Italy
\and
INFN sezione di Milano, Via Celoria 16, 20133 Milano, Italy
\and
Departamento de Astrofísica, Universidad de La Laguna, Ave. Astrofísico Francisco Sánchez, s/n, San Crist\'obal de La Laguna 38206, S/C de Tenerife, Spain
\and
Instituto de Astrofísica de Canarias (IAC), Vía Láctea s/n, San Crist\'obal de La Laguna 38205, S/C de Tenerife, Spain
}

\date{Received XXX; accepted XXX}

\abstract
{Accreting neutron stars in low-mass X-ray binaries allow us to probe the accretion histories that produce recycled millisecond pulsars. A broad distribution of lower spin frequencies together with a narrow group near 575~Hz was identified in a previous work, suggesting that these groups might correspond to two sub-populations of accreting neutron stars.}
{We expand the analysis of the spin distribution with an enlarged sample of 39 accreting neutron stars and investigate whether the high-frequency objects preferentially occur in binaries with longer orbital periods and have higher quiescent surface temperatures. We also ask what the different widths of the spin components imply for the long-term accretion torques of these systems.}
{We compare several empirical models for the spin distribution. Using the preferred model, we compare the orbital periods and quiescent temperatures of the fast and slow neutron stars. We then use stochastic drift--diffusion estimates and simple braking models to determine whether short-timescale torque fluctuations or secular spin-down can account for the observed population widths.}
{The best statistical description of the spin distribution consists of a broad lower-frequency component and a narrow high-frequency concentration centred at $\simeq566$~Hz with a width of $\simeq32$~Hz, consistent with what was previously found. Fast rotators preferentially occupy longer orbital-period binaries, with possibly higher neutron star temperatures.
Short-timescale stochastic accretion-torque fluctuations cannot produce the observed spin-frequency dispersion in either the fast or the slow component, even under generous assumptions.
We show that a non-linear deterministic braking mechanism can instead shift and compress a spin distribution, but the spin data alone do not identify the braking mechanism.}
{The updated sample confirms a narrow high-spin concentration associated with longer orbital periods, with tentative evidence of higher quiescent temperatures. The population widths appear to reflect differences in long-term torque conditions and/or evolutionary histories rather than short-timescale torque noise. If the fast systems are close to spin equilibrium, their narrow distribution requires the combination of physical parameters setting the equilibrium frequency to vary relatively little from source to source. Magnetospheric regulation, correlated variations in accretion rate and magnetic field, and additional spin-down torques, e.g., gravitational-wave emission, remain viable possibilities. Any model that explains the spin distribution should therefore also explain why the fast systems tend to have longer orbital periods and, possibly, higher temperatures.}

\keywords{accretion, accretion disks -- stars: neutron -- X-rays: binaries -- pulsars: general}

\titlerunning{Fast, Long-Period, Hot Accreting Neutron Stars}
\authorrunning{Patruno et al.}

\maketitle

\section{Introduction}\label{sec:intro}

Accreting neutron stars (NSs) in low-mass X-ray binaries (LMXBs) allow us to study how millisecond pulsars are formed. Matter transferred from a low-mass companion can spin the neutron star up to millisecond periods \citep{Alpar1982,RadhakrishnanSrinivasan1982,BhattacharyaVanDenHeuvel1991,Patruno2021}. The spin
frequency reached by the neutron star depends on the net amount of angular momentum transferred during this accretion phase. The final distribution of spin frequencies therefore can be used as a fingerprint to test accretion-torque models and accretion histories~\citep{2017ApJ...850..106P,niang2026}.

An open related question is why accreting NSs do not appear to approach their centrifugal break-up frequencies. Early studies found a possible upper cutoff at around 700\,Hz \citep{Chakrabarty2003,Patruno2010,Patruno2012,Papitto2014}, which is, however, well below the break-up limits of any realistic NS equation of state \citep{Cook1994,Ozel2009,LattimerPrakash2007,HaskellZdunik18}. Several mechanisms could limit further spin-up, like the interaction between the accretion disc and the magnetosphere \citep{GhoshLamb1979,Wang1995,WhiteZhang1997,DaiLi2006,Ho2014}.
Propeller torques, a trapped disc, or the detachment of the companion from its Roche lobe can also remove angular momentum \citep{DAngeloSpruit2010,Dangelo2012,Tauris2012,Patruno2016}. Gravitational waves (GWs), emitted by a deformed star (quadrupolar asymmetry) or through an unstable $r$-mode, offer another possibility
\citep{Bildsten1998,MelatosPayne2005,Haskell2008,Owen1998,Andersson1998,AnderssonKokkotasSchutz1999}.
\citet{2017ApJ...850..106P} studied 29 accreting NSs and found evidence for a broad lower-frequency component and a much narrower component near 575\,Hz. Their two-Gaussian fit gave a mean of 575\,Hz and a standard deviation of 30\,Hz for the fast component. They also estimated a membership-based division point near 540\,Hz, defined as the frequency where the probability of belonging to the fast subpopulation Gaussian is 95\%. The concentration of sources in a narrow range of $\sim500-600$ Hz raised the question about the possible presence of a spin-down torque becoming important only at high frequencies, once the neutron star enters this narrow spin frequency window. Today we know 39 accreting NSs with a securely measured spin frequency. This larger sample, albeit still small, allows us to revisit that picture. We can also ask whether the fast systems differ from the slower ones in their binary properties, and the orbital period provides a simple way to test this. In fact, the donor and the mechanisms that remove orbital angular momentum affect, in principle, the rate and duration of mass transfer and therefore can be used as a simple framework to interpret the observations
\citep{Deloye2003,NelemansJonker2010,Podsiadlowski2002,Knigge2011,Echeveste2024,YangLi2024}. It is therefore reasonable to ask whether systems with different orbital periods have followed different evolutionary and accretion histories. The relation, however, is complicated by the presence of donor and disc irradiation, outflows, or other subtle effects that can alter the evolution~\citep{Patruno2017c,riverasandoval2020}. The transient behaviour of many LMXBs also changes the time-averaged torque relative to that inferred from a single accretion episode
\citep{Bhattacharyya2017,2019MNRAS.488...99G}. 
Therefore, disentangling all these effects may not currently be possible given the small number of known objects. However, we can directly test correlations between NS spin frequencies and orbital parameters.

Additional information may come from the quiescent thermal emission of the accreting NSs. Accreted matter heats the crust, while photons and neutrinos cool down the neutron star by carrying energy away. The surface temperature consequently depends on the long-term accretion rate, but also on the efficiency of cooling and the composition of the NS envelope
\citep{Brown1998,Rutledge2002,Yakovlev2004,Wijnands2013}. A temperature difference between fast and slow systems could therefore suggest different accretion histories for the NSs.

In this paper, we first reassess the spin distribution of accreting neutron stars (Sect.~\ref{sec:spin_distribution}). We then compare the orbital periods (Sect.~\ref{sec:spin_orbit}) and temperatures (Sect.~\ref{sec:temperature}) to the spin distribution to identify potential differences between slow and fast systems. Finally, we ask whether short-lived torque fluctuations can account for the fitted spin distribution widths, and whether simple secular spin-down prescriptions can produce a narrow distribution when the NSs are not in spin equilibrium (Sect.~\ref{sec:torques}). These calculations allow us to discuss what the observations imply for spin evolution (Sect.~\ref{sec:discussion}). This is particularly important for current and future GW searches, as ground-based detectors in the LIGO-Virgo-KAGRA (LVK) network
\citep{LIGOScientific:2014pky,VIRGO:2014yos,KAGRA:2020tym}
have reached the sensitivity where they can set constraints on the scenario in which GW torques are setting the spin equilibrium of these systems \citep[e.g.][]{LIGOScientific:2026ppd, Whelan:2023bha, HaskellRev, 2023APh...15302880W, PhysRevD.105.022002, 2022ApJ...941L..30A, 2020PhRvD.102b3006M, 2017PhRvD..95d2005M, 2015PhRvD..92b3006M}. It is therefore crucial to understand whether GWs are expected in some of these systems and, if so, which objects are more likely to emit them. 

\section{Spin distributions of accreting NSs}\label{sec:distribution}
Our sample contains the 29 systems studied by \citet{2017ApJ...850..106P} and ten additions, listed
in Table~\ref{tab:spin_table} in the appendix. We distinguish accreting millisecond X-ray pulsars (AMXPs), in which accretion-powered pulsations have been reported, from nuclear-powered pulsars (NXPs), whose spins are known only from burst oscillations. A system that shows both kinds of pulsation is classified here as an AMXP. We also classify the three transitional millisecond pulsars as AMXPs, although it is important to keep in mind that their nature might differ from that of the other systems \citep{Papitto2022}. 

\subsection{Spin distribution}\label{sec:spin_distribution}
  
Given the small sample size and the broad, nearly featureless shape of the previously identified slow component~\citep{2017ApJ...850..106P}, we tested a set of simple empirical models to characterize the spin-frequency distribution, and we deliberately kept the number of fitted parameters small. 
We do this to verify
whether the results in our previous work are confirmed, and whether the larger number of objects allows us to constrain the spin distribution further. 
We consider five descriptions: a uniform distribution on the fixed interval $[100,700]$\,Hz, a Gaussian, a skew-normal distribution, a tapered-uniform plus Gaussian mixture, and a two-Gaussian mixture model. The skewed Gaussian is used because it allows a single component to be asymmetric. The two mixture models instead allow a broad distribution at lower spins, together with a concentration at higher frequencies. 

The tapered-uniform plus Gaussian model is
\begin{equation}
 p_{\rm TUG}(\nu)=(1-\lambda_{\rm G})p_{\rm tap}(\nu\mid a,c,w)
 +\lambda_{\rm G}p_{\rm G}(\nu\mid\mu_{\rm G},\sigma_{\rm G}).
 \label{eq:tug}
\end{equation}
Here $p_{\rm G}$ is a Gaussian with mean $\mu_{\rm G}$ and standard deviation $\sigma_{\rm G}$, and $\lambda_{\rm G}$ is its mixture weight. The other component is constant from $a$ to $c-w$, then smoothly decreases to zero between $c-w$ and $c$ through a raised-cosine taper. Each component is normalized to unit integral. We fix the lower edge at $a=100$\,Hz and the taper width at $w=30$\,Hz. The reason for this is that $a$ is a boundary determined by our definition of what an AMXP/NXP is, whereas the taper width is fixed in order to keep the number of free parameters small. Keeping it as a free parameter indeed does not lead to a meaningful fit, as the number of objects is too small. There are therefore four fitted parameters: $\lambda_{\rm G}$, $\mu_{\rm G}$, $\sigma_{\rm G}$ and the endpoint $c$.

For comparison with the earlier work, we also fit a Gaussian mixture model:
\begin{equation}
 p_{2{\rm G}}(\nu)=\lambda p_{\rm G}(\nu\mid\mu_1,\sigma_1)
 +(1-\lambda)p_{\rm G}(\nu\mid\mu_2,\sigma_2),
 \label{eq:two_gaussians}
\end{equation}
with $\mu_1<\mu_2$. This model has five free parameters. Unlike the tapered component, the lower-frequency Gaussian 
does not end at a particular spin, and its tail overlaps the upper component.

We search over the endpoint and use multiple starting points when fitting the mixtures. A Gaussian can otherwise become arbitrarily narrow around one or a few sources, producing a formally good likelihood without a useful description of the population. To prevent this, we require an expected membership of at least 4 objects in each component and a Gaussian width $\sigma\geq5$\,Hz. We also require the two Gaussian means to differ by at least 10\,Hz. In the tapered model, the Gaussian mean must lie at least 10\,Hz above $c$, with $131\leq c\leq689$\,Hz and $\mu_{\rm G}\leq700$\,Hz. We use the same restrictions in the uncertainty calculations below. To compare the fits, we use the small-sample corrected Akaike Information Criterion, AIC$_{\rm c}$~\citep{sugiura1978}. It balances the improvement in likelihood against the additional fitted parameters, with a lower value preferred over a higher one. The comparison is summarized in Table~\ref{tab:models} (Appendix). We use the AIC$_{\rm c}$ to rank these empirical descriptions (which is different from assigning probabilities to physical populations). 

The tapered-uniform plus Gaussian model has the lowest AIC$_{\rm c}$. It places the upper concentration at $\mu_{\rm G}=565.7$\,Hz with $\sigma_{\rm G}=32.4$\,Hz. The
two-Gaussian fit gives closely similar values, $\mu_2=568.3$\,Hz and
$\sigma_2=31.4$\,Hz, and is worse by $\Delta{\rm AIC}_{\rm c}=5.50$. Both fits
are consistent with the location and width reported by \citet{2017ApJ...850..106P}. However, the skew-normal is only $\Delta{\rm AIC}_{\rm c}=3.17$ above the preferred model when a wide range of shape parameters is allowed. In that limit it approaches a distribution with a sharp edge. Thus, a single asymmetric component remains an alternative, with the concentration at high spin frequencies.

We estimate parameter uncertainties by drawing 2000 synthetic datasets from each fitted mixture and refitting them with the same restrictions. Each dataset contains 39 spins. Table~\ref{tab:model_params} in Appendix~\ref{app:mixture_par} reports the central 95\% percentile intervals. The tapered endpoint is $c=487.5$\,Hz, with an interval $[385.9,579.6]$\,Hz. These intervals describe uncertainty within each assumed model.

\subsubsection{Where should the fast and slow systems be divided?}
\label{sec:spin_boundary}

\begin{figure*}
\centering
\includegraphics[width=\textwidth]{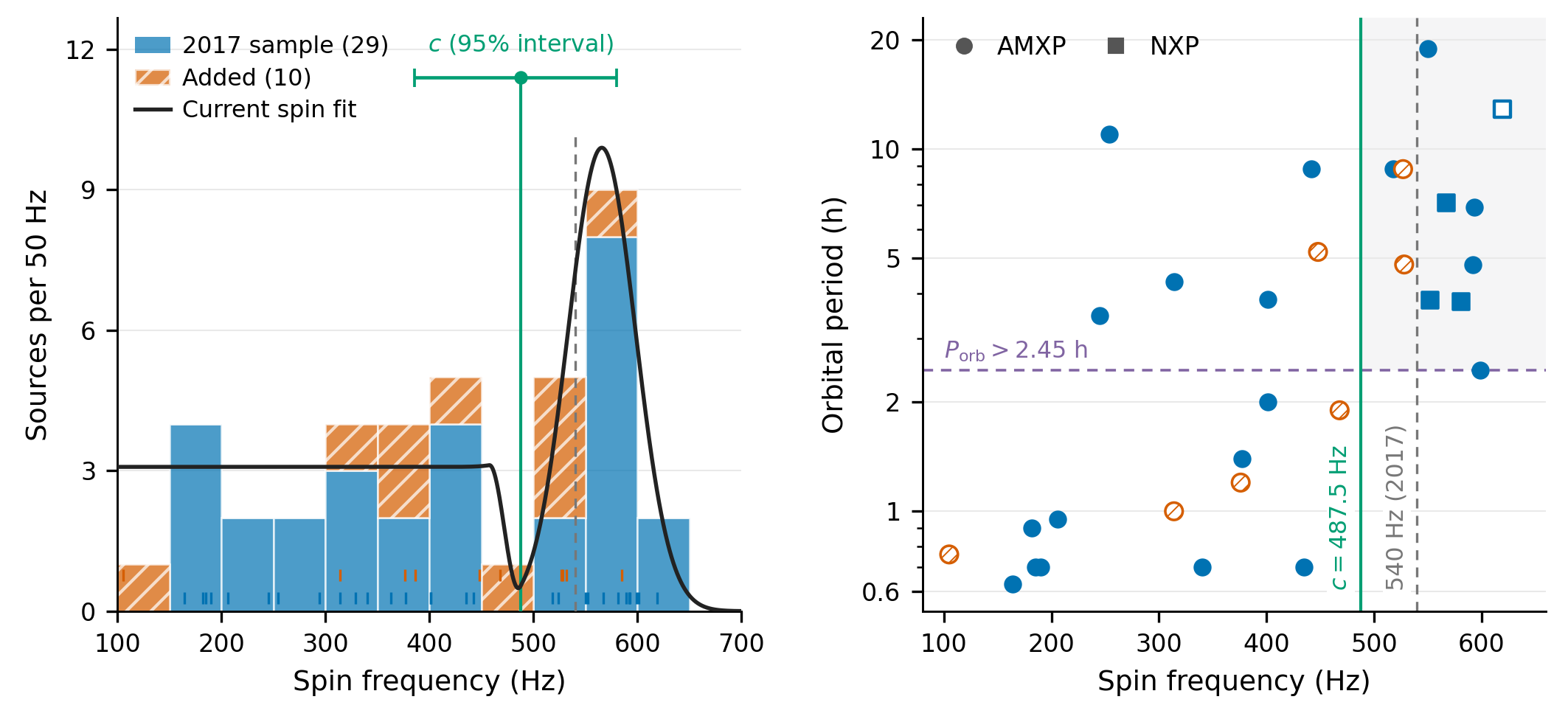}
\caption{Spin frequencies and orbital periods of the accreting NS sample. Left: the histogram separates the 29 members of the 2017 sample (blue) from the ten additions (orange hatching). Short marks along the axis show the individual spins, and the black curve is the fitted tapered-uniform plus Gaussian density scaled to the number of sources per bin. The green line marks the current
tapered endpoint, with the horizontal bar giving the 95\% bootstrap interval. The grey dashed line is the 540\,Hz classification cut from \citet{2017ApJ...850..106P}, which uses a different definition (Sect.~\ref{sec:spin_boundary}). Right: orbital period vs spin frequency. Colours have the same meaning as in the left panel, with filled and open hatched symbols distinguishing historical and added sources, while circles and squares denote AMXPs and NXPs. 4U 1608--52 is shown as a blue open square for reference, but is not included in the spin-orbit analysis, for conservative reasons, as explained in the text. The current division gives 10 fast (not counting 4U 1608--52) and 19 slow systems. Fast systems tend to have longer periods, with two added sources above the new division.}
\label{fig:spin_orbit}
\label{fig:spin_cumulative}
\end{figure*}

The new endpoint $c$ cannot be compared directly with the old 540\,Hz cutoff value \citep{2017ApJ...850..106P} as if both measured the same boundary. The reason is that while the tapered model uses the parameter $c$ as the lower-frequency component vanishing point, the 2017
two-Gaussian model had no corresponding parameter. In other words, a source was assigned to the fast group once its fitted probability of belonging to the slow Gaussian fell to 0.05. This gave a cut at 538\,Hz, with a quoted 95\% interval of 526--548\,Hz \citep{2017ApJ...850..106P}. For the mixture in Eq.~\eqref{eq:two_gaussians}, the fitted membership probability of the upper component is
\begin{equation}
 P_{\rm fast}(\nu)=
 \frac{(1-\lambda)p_{\rm G}(\nu\mid\mu_2,\sigma_2)}
 {p_{2{\rm G}}(\nu)}.
 \label{eq:membership}
\end{equation}
We can therefore apply the same $P_{\rm fast}=0.95$ rule to the current fit, which gives a lower crossing at $569.7$\,Hz. 
However, it is important to notice that in 579 of the 2000 refits (29\%), $P_{\rm fast}$ never reaches 0.95. In those cases the two fitted components overlap too strongly for any frequency to
meet the rule. Furthermore, when the Gaussian widths differ, the broader component can dominate again at sufficiently high frequency. A lower crossing is consequently not necessarily a cutoff above which every source has a high fast-component membership probability. A division at equal membership, $P_{\rm fast}=0.5$, occurs at $506.4$\,Hz on the lower side of the concentration. The position of the high-frequency concentration is relatively well constrained, with both mixture fits placing it near 566--568 Hz, whereas the frequency that separates it from the slower systems is much less certain. In this updated work, we use the endpoint $c$ from our preferred fit as the working division in the following comparisons. The fitted spin distribution and the secure orbital-period sample are shown in Fig.~\ref{fig:spin_cumulative}.

\subsection{Spin--orbit correlation}\label{sec:spin_orbit}
If the two spin regimes correspond to different binary evolution channels, they may also differ in orbital period. To test this possibility, we investigated whether the fast NSs preferentially occupy longer orbital periods than the slow ones. Of the 39 systems with measured spin frequencies, 29 have reliable orbital–period estimates, plus one object (4U 1608--52, \citealt{Wachter2002}) which has an orbital period determined from superhump observations (Table~\ref{tab:spin_table}). We exclude this indirect orbital-period estimate from the analysis because it has not been confirmed by a direct orbital measurement. This choice is conservative because 4U 1608–52 spins at 619 Hz and has a proposed long orbital period of about 12.9 h. Including it increases the effect size (see e.g., Table D.1) and strengthens the conclusions of this paper. 
In our analysis, we quantify the strength of the effect as
\begin{equation}
\widehat A=\frac{U_{\rm fast}}{n_{\rm F}n_{\rm S}},
\end{equation}
where $n_{\rm F}$ and $n_{\rm S}$ are the numbers of systems in the fast and slow groups, respectively, and $U_{\rm fast}$ is the Mann--Whitney statistic computed for the fast group~\citep{mann1947}. The quantity $\widehat A$ is the probability that a randomly selected fast system has a longer orbital period than a randomly selected slow system, where $U_{\rm fast}$ counts the number of fast--slow pairs satisfying this ordering. A value of $\widehat A=0.5$ would indicate no systematic ordering between the two groups, whereas $\widehat A>0.5$ indicates a tendency for the fast systems to have longer orbital periods.
For the current sample, containing 10 fast and 19 slow systems, we find $U_{\rm fast}=161$, corresponding to $\widehat A=0.847$ and a one-sided probability of $p=0.00078$. Thus, approximately 85\% of all fast--slow pairs are ordered such that the faster-spinning system also has the longer orbital period. 
We also repeated the comparison using only AMXPs and obtained $\widehat A=0.872$ and $p=0.001307$ (see Table~\ref{tab:orbit_comparisons}). The same ordering is therefore present within the AMXP population itself, indicating that the result is not simply produced by combining different classes of systems. Furthermore, defining the fast and slow populations using the equal-membership division of the two-Gaussian model leads to the same group assignments and hence the results remain unchanged.

We then assessed sensitivity to the adopted spin division at the thresholds listed in Table~\ref{tab:cutoff_sensitivity}. The observed tendency towards longer orbital periods in the fast group persists at all tabulated thresholds. However, $\widehat{A}$ decreases from 0.89 at 436 Hz to 0.69 at 556 Hz, while the corresponding one-sided permutation probability increases from $7.4\times10^{-5}$ to $0.1$. We also repeated the calculation by propagating the uncertainty on $c$, i.e., we used all 2000 fitted endpoints, reclassified the 29 systems with orbital periods into fast and slow groups, and repeated the orbital comparison. We obtain 1989 non-empty comparisons, of which 99.7\% show the same longer-period ordering. In 91.4\% of these realizations, the one-sided p-value is $<0.05$. The orbital-period result is therefore not driven by our choice of 487.5 Hz as the dividing frequency. The same ordering is found over almost the entire bootstrap range of the boundary. 

\subsection{Temperature--spin correlation analysis}\label{sec:temperature}
\begin{figure}
\centering
\includegraphics[width=\columnwidth]{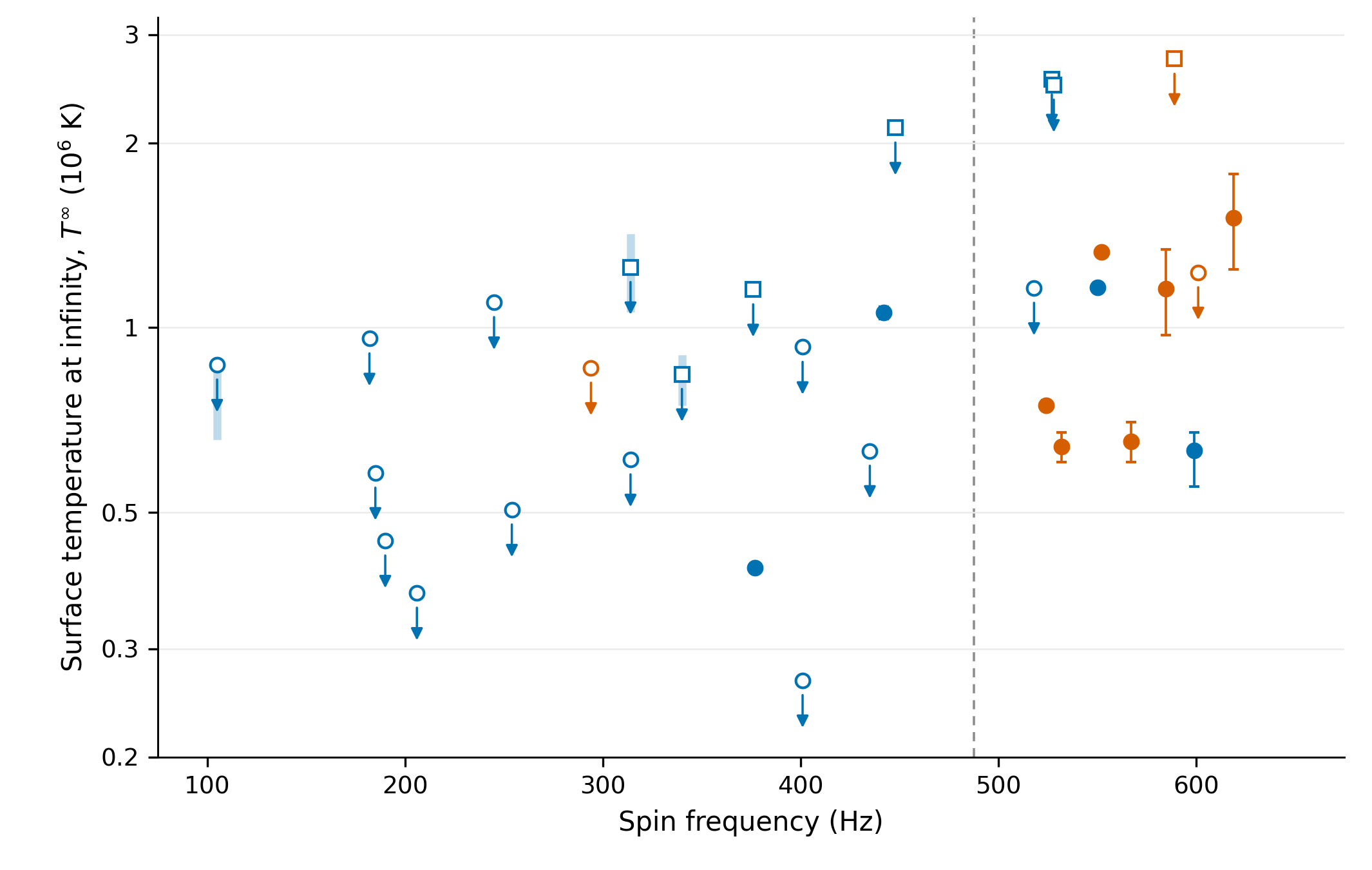}
\caption{Spin frequency and quiescent thermal constraints for all
31 sources with a temperature estimate. Blue and orange indicate AMXPs and NXPs, respectively. Filled circles show atmosphere-based measurements; open circles and squares show atmosphere-based and blackbody-derived upper limits, respectively. Downward arrows indicate upper limits. Error bars show reported uncertainties. Pale vertical segments indicate ranges associated with uncertainties in distance or redshift assumptions. The dashed line marks the 487.5~Hz endpoint fitted to spins alone. References are
in Table~\ref{tab:thermal_all}.}
\label{fig:temperature_updated}
\end{figure}

We next examined whether the fast and slow-spinning systems have different quiescent surface-temperature distributions. This comparison is complicated by the fact that many sources have only upper limits on their temperature. We therefore choose a method of analysis that could correctly account for this fact. We begin with all 20 sources in the surface-temperature compilation of
\citet{gusakov2014} that occur in our 39-source dataset, and add
11 sources with published quiescent measurements. The resulting
comparison contains 31 nominal whole-surface constraints:
24 atmosphere estimates or limits and 7 approximate blackbody upper limits. When a thermal spectral component has a surface area much smaller than the NS surface, as expected, for example, from local hot spot emission, we do not include that source in our sample. The same is done when the temperature might have an ambiguous surface/disc origin. The final list of selected sources is in Table~\ref{tab:thermal_all}. The resulting constraints are shown in Fig.~\ref{fig:temperature_updated}.

We express the temperatures as seen by a distant observer, denoted by $T^\infty$, in millions of kelvin (MK). For an atmosphere fit reporting a local surface temperature $T_{\rm s}$, the conversion is
\begin{equation}
 T^\infty=gT_{\rm s},\qquad
 g=\left(1-\frac{2GM}{Rc^2}\right)^{1/2}.
 \label{eq:thermal_frame}
\end{equation}
Here $M$ and $R$ are the NS mass and radius assumed in that fit,
$G$ is Newton's gravitational constant and $c$ is the speed of light. The dimensionless factor $g$ describes gravitational redshift. Quiescent measurements are used for EXO~0748--676, MXB~1659--298 and Aql~X--1
\citep{thermalParikh2020,thermalParikh2019,thermalWaterhouse2016}.
For HETE~J1900.1--2455 we use the representative
$k_{\rm B}T^\infty\simeq35$~eV estimate adopted by
\citet{thermalDegenaar2021}, corresponding to about $0.41$~MK in 2018. Their approximately $30$--$39$~eV range reflects the dependence of their upper temperature constraints on mass, radius and distance. We assume that no residual accretion was occurring during the observations. Of course, if this assumption is lifted, the temperatures cannot be considered reliable proxies of the NS temperature. 

We also use quiescent luminosities to infer upper limits on the temperature, if 
one was not explicitly reported in the literature. For a bolometric luminosity limit $L^\infty_{\rm lim}$, we assign all allowed emission to a uniformly emitting blackbody surface giving:
\begin{equation}
 T^\infty_{\rm BB,lim}=
 \left[\frac{L^\infty_{\rm lim}}
 {4\pi(R^\infty)^2\sigma_{\rm SB}}\right]^{1/4},
 \qquad R^\infty=R/g .
 \label{eq:thermal_bolometric}
\end{equation}
The subscript BB denotes the blackbody approximation, $R^\infty$ is the
apparent stellar radius, and $\sigma_{\rm SB}$ is the Stefan--Boltzmann
constant. Some flux-derived entries instead refer to finite X-ray bands, so we solve
\begin{equation}
 F_{[E_1,E_2]}=
 \left(\frac{R^\infty}{d}\right)^2\sigma_{\rm SB}(T^\infty_{\rm BB})^4
 \frac{15}{\pi^4}
 \int_{E_1/(k_{\rm B}T^\infty_{\rm BB})}^{E_2/(k_{\rm B}T^\infty_{\rm BB})}
 \frac{x^3 A(xk_{\rm B}T^\infty_{\rm BB})}{e^x-1}\,dx .
 \label{eq:thermal_band}
\end{equation}
Here $F_{[E_1,E_2]}$ is the reported energy flux between photon energies
$E_1$ and $E_2$, $d$ is the distance, and
$x=E/(k_{\rm B}T^\infty_{\rm BB})$ is dimensionless photon energy.
The factor $A(E)$ is the fraction of radiation transmitted through intervening material, equal to one for an absorption-corrected flux. For the absorbed flux limits of GRS~1741.9--2853 and MAXI~J1816--195, we include approximate photoelectric attenuation using their reported hydrogen column density
$N_{\rm H}$ and the cross sections of \citet{thermalMorrison1983}. For GRS~1741 the calculation also includes the dust-scattering prescription described by
\citet{thermalMuno2003}. For these seven calculations only, we assume $M=1.4$ solar masses and $R=10$~km, giving $R^\infty=13.06$~km. For example, the quiescent 0.5--10~keV flux limit of IGR~J17494--3030 gives
$T^\infty_{\rm BB}\lesssim1.15$~MK at 8~kpc
\citep{ArmasPadilla2013_17494}, while the upper limit for
IGR~J17591--2342 gives approximately $2.53$~MK
\citep{Gusinskaia2020_17591}.   

Since many of the temperatures reported in the literature are upper limits, we selected a two-sided imputation--permutation log-rank test \citep{WangLagakosGray2010}, which was developed for small samples and allows the two groups to have different patterns of censoring. Here the temperatures are left-censored, i.e., for a source with $T^\infty$ below the measured upper limit, the true temperature is unknown but constrained to lie below the upper limit. The analysis includes all 31 temperature constraints: 10 measurements and 21 upper limits. Although the atmosphere-based measurements and the blackbody-derived limits rely on different spectral assumptions, we treat both as approximate constraints on the quiescent surface temperature. We divide the systems according to the spin boundary obtained from the spin-distribution analysis, $\nu_{\rm c}=487.5$~Hz. Systems with $\nu>\nu_{\rm c}$ are assigned to the fast group and those with $\nu\leq\nu_{\rm c}$ to the slow group. This gives 13 fast systems, of which 5 have temperature upper limits, and 18 slow systems, of which 16 have upper limits.
To include these upper limits without assigning arbitrary temperatures to them, we reverse the temperature axis and apply a survival-analysis method, i.e., each upper limit represents a source that lies somewhere below its reported temperature limit. The log-rank statistic then asks whether the ordering of the temperature constraints differs systematically between the fast and slow groups, while taking into account that some sources are censored rather than measured. To determine the significance of our results, we first generate $10^6$ realizations compatible with the measured temperatures and upper limits. Under the null hypothesis that the fast and slow NSs are drawn from the same underlying temperature distribution, the inferred temperatures are reassigned between the two groups and the appropriate censoring pattern is then reapplied. In this way we obtain two-sided $p$-values that give the probability, under the null hypothesis, of obtaining a separation at least as strong as the observed one. 
Two-sided means that a difference favouring either hotter fast systems or hotter slow systems counts as evidence against the null hypothesis. We find that the fast-spinning systems tend to have higher quiescent temperatures with $p=8.6\times10^{-5}$.
It is important to note, however, that the fast and slow groups contain different proportions of AMXPs and NXPs. Even if we do not know whether such distinction matters from an evolutionary point of view (and thus for the temperature), we repeated the comparison within each source class and combined the resulting class-specific statistics. The temperature difference remains present, with
$p=0.015$. The observed ordering is therefore not explained solely by the different mixture of source classes, although the class-adjusted comparison is necessarily less significant than the unadjusted one given the smaller number of objects. Restricting the same thermal comparison to the 22 AMXPs also shows the same hotter-fast ordering (five fast and 17 slow systems, with two-sided $p=0.026$). As a sensitivity test, we also repeated the analysis using the historical fast vs slow division at 540~Hz adopted in \citet{2017ApJ...850..106P}. We obtain $p=6.3\times10^{-4}$ for the full sample and $p=0.0053$ after the same source-class adjustment. The qualitative result therefore does not depend on using the newly fitted value of $\nu_{\rm c}$. 

It is important to stress that several limitations apply when interpreting these probabilities. First, the small sample size means that some simulations in which the two groups were drawn from the same temperature distribution produced rejection rates between $3.9$ and $6.2$ per cent for a nominal $5$ per cent test. Second, propagating the uncertainty in $c$, i.e., using all 2000 fitted endpoints, preserves the hotter-fast log-rank ordering in all 1999 non-empty comparisons, with the caveat that at the 97.5th-percentile endpoint the conditional two-sided probability increases to $p=0.054$. Third, the method assumes that, within the modelled groups, the censoring process is sufficiently independent of the underlying temperature. In practice, whether a temperature is measured or only constrained by an upper limit can depend on factors like the source distance, absorption, exposure, and observing sensitivity. Given these limitations, the data suggest that the fast systems are hotter, but we do not consider this result conclusive.

\section{Phenomenological spin-equilibrium interpretation}\label{sec:torques}
The empirical analysis above identifies a broad population of slowly rotating systems and a much narrower concentration at high spin frequency, centred at about 566~Hz with a width of about 32~Hz. The fast systems also tend to have longer orbital periods and higher quiescent temperatures. The statistical analysis alone, however, does not tell us why the two spin components have such different widths. A natural possibility is that the systems are close to a long-term spin equilibrium. In this picture, the spin of each neutron star fluctuates around an equilibrium value set by the balance between spin-up and spin-down torques. There are then two ways to produce the observed width. Individual stars may wander around their own equilibrium frequency because the accretion torque varies with time, or different stars may have different equilibrium frequencies because their long-term accretion rates, magnetic fields, duty cycles, or other physical properties differ. We first ask whether the first effect alone can be large enough to explain the observed widths. 

Consider a source $i$ with mean spin derivative $A_i(\nu)$ and a stable equilibrium at frequency $\mu_i$. At equilibrium, the mean torque vanishes and a small displacement from equilibrium must produce a restoring torque, so that
\begin{equation}
A_i(\mu_i)=0,
\qquad
A_i'(\mu_i)<0 .
\end{equation}
Close to $\mu_i$, the torque can therefore be linearized as
$A_i(\nu)\simeq-\kappa_i(\nu-\mu_i)$, where
$\kappa_i=-A_i'(\mu_i)>0$ measures how strongly the system is driven back
towards equilibrium.

We describe the short-term torque fluctuations in terms of a phenomenological mean-reverting stochastic process, as used also by \citet{Melatos23} and \citet{Melatos26} to model hidden accretion variables and spin wandering in accreting NSs. We therefore describe the local spin evolution as:
\begin{equation}
d\nu_i =
-\kappa_i(\nu_i-\mu_i)\,dt
+
\sqrt{2D_{\nu,i}}\,dW_i ,
\end{equation}
where $D_{\nu,i}$ is the spin-frequency diffusion coefficient and $dW_i$ is a Wiener increment. This is simply a local description of a star whose spin is continually perturbed by torque fluctuations while the mean torque tends to restore it towards $\mu_i$. If the star starts at its equilibrium frequency, the variance produced by these fluctuations after a time $\tau$ is
\begin{equation}
\sigma^2_{\rm stoch,i}(\tau)
=
\frac{D_{\nu,i}}{\kappa_i}
\left(1-e^{-2\kappa_i \tau}\right)
\leq 2D_{\nu,i}\tau .
\end{equation}
The important point is that the stochastic broadening does not grow without limit. At early times, it approaches the free-diffusion result, whereas at late times the restoring torque limits the variance to $D_{\nu,i}/\kappa_i$.

For a population of sources close to equilibrium, the total observed spin variance can therefore be written as
\begin{equation}
{\rm Var}(\nu)
=
{\rm Var}(\mu_i)
+
\left\langle
\frac{D_{\nu,i}}{\kappa_i}
\right\rangle .
\end{equation}
The first term describes real differences in equilibrium spin frequency from one system to another. The second describes the additional width produced by each neutron star wandering around its own equilibrium. We therefore investigate whether this second term can plausibly account for the observed width of either spin component.

To estimate its magnitude, we relate $D_\nu$ to the fluctuating spin derivative. We define $\delta\dot{\nu}$ as the zero-mean part of the spin derivative, with rms amplitude $\sigma_{\dot{\nu}}$, and
\begin{equation}
C_{\dot{\nu}}(u)
=
\left\langle
\delta\dot{\nu}(t)
\delta\dot{\nu}(t+u)
\right\rangle
\end{equation}
its autocovariance at time lag $u$. We define the integral correlation time as
\begin{equation}
\tau_{\rm int}
=
\frac{1}{\sigma_{\dot{\nu}}^2}
\int_0^\infty C_{\dot{\nu}}(u)\,du,
\qquad
D_\nu
=
\sigma_{\dot{\nu}}^2\tau_{\rm int}.
\end{equation}
Thus, $\tau_{\rm int}$ measures for how long a typical torque fluctuation
remains correlated. For times much longer than this correlation time, the
corresponding free-diffusion displacement is
\begin{equation}
\Delta\nu_{\rm diff}
\simeq
\sqrt{2D_\nu \tau}
=
\sigma_{\dot{\nu}}\sqrt{2\tau_{\rm int}\tau}.
\end{equation}
Numerically,
\begin{equation}\label{eq:delta_nu}
\Delta\nu_{\rm diff}
\simeq
7.9~{\rm Hz}
\left(
\frac{\sigma_{\dot{\nu}}}
{10^{-11}\ {\rm Hz\,s^{-1}}}
\right)
\left(
\frac{\tau_{\rm int}}
{10^7\ {\rm s}}
\right)^{1/2}
\left(
\frac{\tau}
{10^9\ {\rm yr}}
\right)^{1/2}.
\end{equation}

This estimate is conservative, as a correlation time of $10^7$~s is already comparable with the duration of a long accretion episode, while an rms spin derivative of $10^{-11}$~Hz~s$^{-1}$ is larger than the characteristic accretion torque expected for LMXBs. To verify this, we can consider the usual order-of-magnitude accretion torque
\begin{equation}
N_{\rm acc}
\sim
\dot{M}\sqrt{GMR_{\rm m}},
\qquad
\dot{\nu}_{\rm acc}
=
\frac{N_{\rm acc}}{2\pi I},
\end{equation}
where $\dot{M}$ is the mass-accretion rate, $M$ and $I$ are the neutron-star mass and moment of inertia, and $R_{\rm m}$ is the magnetospheric radius at
which the accreting matter transfers its angular momentum to the star. For a steadily accreting system, the magnetospheric radius cannot greatly exceed the corotation radius without changing the nature of the accretion
flow. The corotation radius is
\begin{equation}
R_{\rm co}
=
\left[
\frac{GM}{(2\pi\nu)^2}
\right]^{1/3}.
\end{equation}
For $M=1.4\,M_\odot$, $I=10^{45}$~g~cm$^2$, $\nu=700$~Hz and
$\dot{M}=10^{18}$~g~s$^{-1}$, taking the material lever arm to be
$R_{\rm co}=2.13\times10^6$~cm gives
\begin{equation}
\dot{\nu}_{\rm acc}
\simeq
3.2\times10^{-12}\ {\rm Hz\,s^{-1}} .
\label{eq:max-spin-up}
\end{equation}
The value $\sigma_{\dot{\nu}}=10^{-11}$~Hz~s$^{-1}$ used above allows fluctuations substantially larger than this characteristic accretion-torque scale.

Even with these generous assumptions, the accumulated stochastic displacement over $10^9$~yr is only of order a few Hz. It is worth noticing that empirical constraints on the amplitude and correlation timescale of accretion torque fluctuations are rare, especially for LMXBs. In the High Mass X-ray Binary SXP~18.3 and subsequently in 24 other similar systems in the Small Magellanic Cloud, however, \citet{oleary2024, oleary2024b} found a relaxation timescale of $\sim10^6-10^{8}$~s and typical rms
fluctuations in the mass-accretion rate of order $20\%$ of its mean value, consistent with the fiducial values chosen in this work. We caution that the applicability of this result to AMXPs is not proven since HMXBs behave quite differently. However, given the limited number of observational constraints, it is reassuring to see that the observations support the assumptions adopted in this work. 
The few Hz displacement we find with Eq.~\ref{eq:delta_nu} is therefore far smaller than the roughly $100$--$200$~Hz scale of the broad slow component and is also well below the best-fitting width of the fast component, $\sigma_{\rm F}\simeq32$~Hz. Furthermore, the free-diffusion estimate used here is itself conservative because when a restoring torque is present, the stochastic variance eventually saturates rather than continuing to grow indefinitely as $\tau^{1/2}$. The observed component widths therefore cannot be explained primarily by ordinary short-timescale accretion-torque fluctuations. In a spin equilibrium interpretation, most of the width must instead reflect differences between systems in their long-term equilibrium frequencies. For the slow NS group,
this spread is large, implying substantial source-to-source diversity in the quantities that determine the equilibrium spin. These may involve the long-term accretion rate, magnetic field, duty cycle, donor properties, or the history of propeller phases and outflows. The fast component is much narrower, which instead suggests that the combination of physical parameters setting its characteristic spin is considerably more tightly regulated.

This interpretation does not require every source to be exactly at torque balance at the present epoch. Some systems may still be evolving towards or away from equilibrium. In that case, the observed population width also contains differences in evolutionary stage and long-term torque history.

\subsection{Secular braking versus stochastic diffusion}\label{sec:braking}
The previous section considered stochastic torque fluctuations around a local spin equilibrium and showed that ordinary accretion-driven noise is too weak to account for the observed widths of the spin components. We now ask a different question: how does a coherent long-term spin-down torque change the shape of a spin distribution, and how does this compete with stochastic broadening? The aim here is to build a simple phenomenological calculation to separate the effects of systematic braking from those of stochastic torque fluctuations. The reason why we focus on a spin-down torque only is that we are investigating a scenario where some of the accreting NSs experience a negligible accretion torque, for example because they enter a propeller or a trapped-disc regime, or because a strong braking mechanism like a gravitational wave instability dominates the torque. We therefore want to understand whether the resulting spin-down would be able to explain the observed spin components in the spin distribution.


\begin{figure*}[t]
\centering
\includegraphics[width=\linewidth]{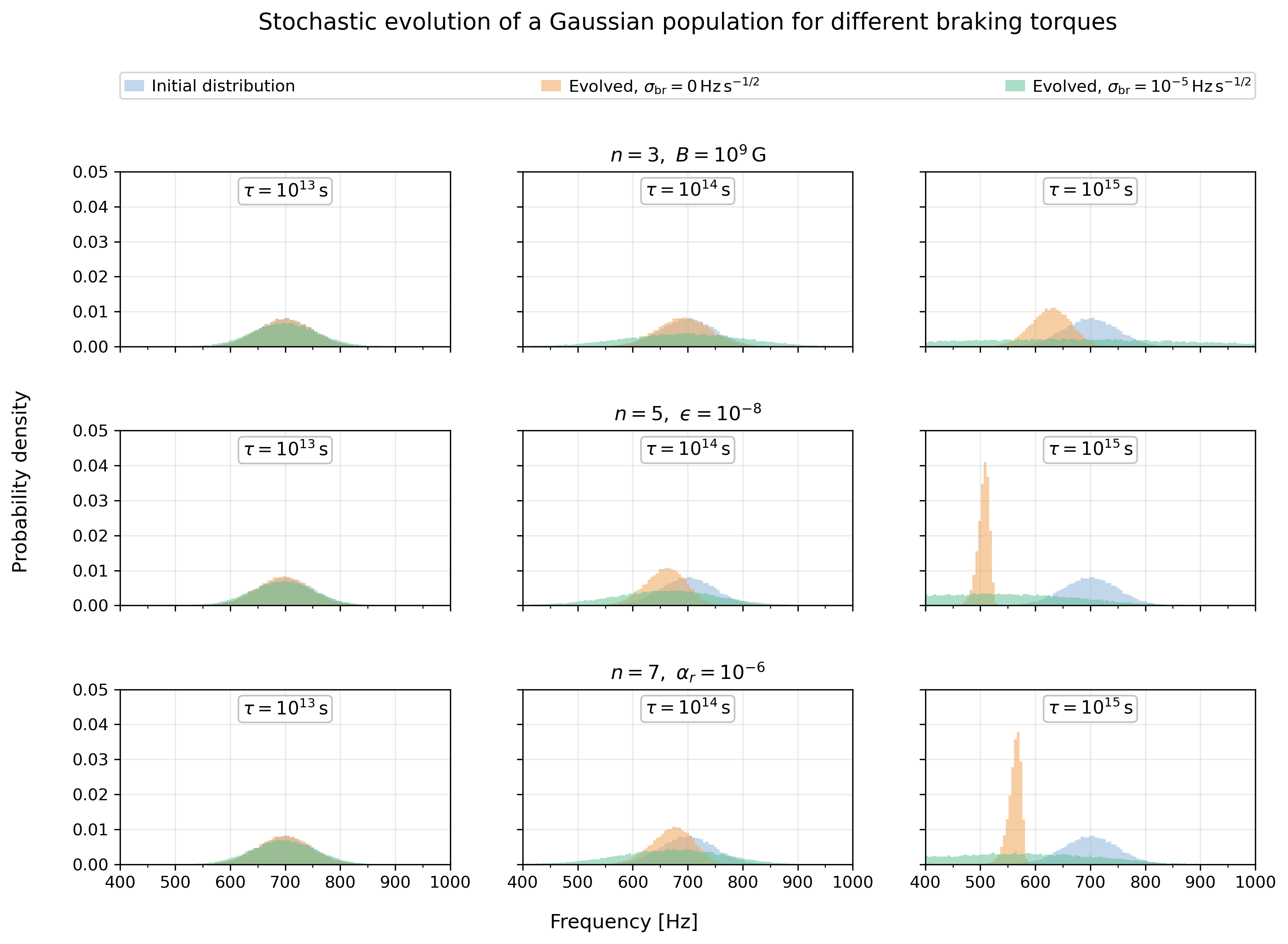}
\caption{
Representative evolution of an initial Gaussian spin distribution under the single-channel stochastic braking model of Eq.~\ref{eq:stochastic_braking}. Columns correspond to increasing
integration time, $\tau$, while rows correspond to different braking
mechanisms. Each row shows the evolution both in the absence of a
stochastic torque, $\sigma_{\rm br}=0 \,\rm{Hz}\,s^{-1/2}$, and for a finite stochastic amplitude $\sigma_{\rm br}=10^{-5} \,\rm{Hz}\,s^{-1/2} $. The figure illustrates the competition between deterministic braking, which shifts and can sharpen the distribution, and stochastic forcing, which broadens and smooths it.  However, the slow-branch-like flat morphologies require values of $\sigma_{\rm br}$ far above those implied by ordinary accretion-torque noise, even under optimistic assumptions for $\sigma_{\dot\nu}$ and $\tau_{\rm int}$. Narrow shifted distributions occur instead when deterministic braking dominates.
}
\label{fig:simulations}
\end{figure*}

We consider an ensemble with three illustrative braking laws: magnetic-dipole-like braking ($n=3$), mass-quadrupole gravitational-wave braking ($n=5$), and $r$-mode-like braking ($n=7$).  The deterministic drift is taken to be
\[
A(\nu)=-a\nu^n ,
\]
so the stochastic evolution is
\begin{equation}
d\nu = -a\,\nu^n\,dt + \sigma_{\rm br}\,dW_t,
\label{eq:stochastic_braking}
\end{equation}
where $a>0$ sets the braking strength, $n$ is the braking index, and
$\sigma_{\rm br}=\sqrt{2D_\nu}$ is a phenomenological white-noise broadening amplitude with units ${\rm Hz\,s^{-1/2}}$.
Unlike the local equilibrium model of Sect.~\ref{sec:torques}, Eq.~\ref{eq:stochastic_braking} has no finite-frequency attractor. It therefore describes the limiting case in which spin-up is absent (or negligible), as may occur in a low-accretion/propeller-dominated case, or when a hidden braking torque dominates the spin evolution.

The coefficient $a$ depends on the braking mechanism considered. For magnetic-dipole braking ($n=3$), $a \propto B^2$, where $B$ is the magnetic field at the magnetic pole of the NS. For a permanent mass quadrupole ($n=5$), $a \propto \epsilon^2$, where $\epsilon \equiv |I_{xx}-I_{yy}|/I_{zz}$ is the equatorial ellipticity, and $I_{xx}$, $I_{yy}$, and $I_{zz}$ are the principal moments of inertia. For r-mode braking ($n=7$), $a \propto \alpha_{\rm r}^2$, where $\alpha_{\rm r}$ is the saturation amplitude of the mode.

An initial Gaussian distribution is used here only as a representative case that provides a common, moderately narrow initial distribution from which we can examine how deterministic braking and stochastic diffusion alter the population. We choose an initial width of 50 Hz, which is neither too close to the width of the fast component nor to the width of the slow component. At each time, the absolute width of the evolved distribution necessarily depends on this assumed initial distribution. Consequently, we cannot use these calculations to infer a unique value of $\sigma_{\rm br}$ required to reproduce the observed slow or fast branch. Instead, the relevant quantity is the additional broadening produced by the stochastic term relative to the deterministic evolution from the same initial distribution.

We integrated Eq.~\ref{eq:stochastic_braking} with an Euler--Maruyama
scheme for ensembles of systems, imposing a non-negative spin floor.
The simulations show the expected competition between the two terms.
The deterministic braking term shifts systems towards lower spin
frequencies and, because the braking rate increases with spin frequency,
can sharpen or compress the high-frequency part of the distribution.
The stochastic term instead broadens and smooths the distribution.
When stochastic broadening dominates, an initially narrow distribution
can develop a broad morphology, whereas when
deterministic braking dominates, the distribution remains comparatively
narrow but is displaced towards lower frequencies, sometimes producing
a sharp edge or transient pile-up.

In our calculations, we explored a range of stochastic amplitudes
$\sigma_{\rm br}$ and torque strengths corresponding to different
magnetic fields, ellipticities, and $r$-mode saturation amplitudes.
Specifically, we considered
$\sigma_{\rm br}\in[10^{-10},10^{-3}]\,{\rm Hz\,s^{-1/2}}$,
$B\in[10^9,10^{12}]\,{\rm G}$,
$\epsilon\in[10^{-9},10^{-6}]$, and
$\alpha_{\rm r}\in[10^{-8},10^{-1}]$.
This deliberately broad parameter space extends well beyond what is
expected for AMXPs and is intended to identify the qualitative forms of
evolution that are possible and when stochastic broadening begins to
compete with deterministic braking. 
Across the explored parameter space,
stochastic broadening remains negligible for
$\sigma_{\rm br}\lesssim10^{-6}\,{\rm Hz\,s^{-1/2}}$, independently of
the particular braking mechanism. 
Above this threshold, diffusion
progressively broadens and flattens the population, with broad,
slow-branch-like morphologies appearing only for
$\sigma_{\rm br}$ of order $10^{-6}\,{\rm Hz\,s^{-1/2}}$ or larger.
Recent studies of accretion-torque fluctuations
\citep{oleary2024,oleary2024b,antonelliGravitationalPulsarsCorrelations2025}
suggest stochastic amplitudes well below this threshold, typically of
order $\sigma_{\rm br}\sim10^{-8}\,{\rm Hz\,s^{-1/2}}$.
Such fluctuations would therefore have a negligible effect on the
spin distribution.


Figure~\ref{fig:simulations} illustrates this competition for the three
braking indices. For each mechanism, we show the deterministic evolution
and the corresponding evolution with
$\sigma_{\rm br}=10^{-5}\,{\rm Hz\,s^{-1/2}}$.
This value is not intended to represent a realistic level of ordinary
accretion-torque noise, but is chosen to illustrate the morphological
effect of stochastic broadening once it becomes significant.
The comparison illustrates that substantial stochastic broadening requires
noise amplitudes well above those expected from ordinary accretion-torque
fluctuations, 
whereas deterministic braking can significantly reshape the distribution.
We examine the deterministic case in more detail in the following
section and assess the corresponding values of $B$, $\epsilon$, and $\alpha_{\rm r}$.

\subsection{Narrowing under deterministic braking}
We now examine a more specific deterministic case.
We set the stochastic term to zero and fix the braking strength by assuming
that the initial population was in spin equilibrium near $700\,\mathrm{Hz}$,
where the braking torque balances an accretion spin-up torque of order
$10^{-11}\,\mathrm{Hz\,s^{-1}}$, adopted here as a conservative upper limit.
The purpose is to isolate whether nonlinear spin-down can compress an initially
broad ($\sim 50$ Hz) high-frequency population and thereby contribute to the
narrow fast branch.
This determines, for each braking mechanism, the corresponding values of the magnetic field $B$, ellipticity $\epsilon$, or $r$-mode saturation amplitude
$\alpha_{\rm r}$. We then study the subsequent evolution after the spin-up torque ceases, so that the spin evolution is governed only by a braking torque of the form
\begin{equation}
    A(\nu)=-a\nu^n.
\end{equation}
In this case, a population initially distributed as a Gaussian in
spin frequency, with mean $\mu$ and width $\sigma$, evolves after a time $\tau$ into
\begin{equation}
\begin{aligned}
f(\nu)
={}&
\frac{1}{\sigma\sqrt{2\pi}}
\exp\!\left[
-\frac{1}{2\sigma^2}
\left(
\nu
\left[1-(n-1)a\tau\nu^{n-1}\right]^{-1/(n-1)}
-\mu
\right)^2
\right]
\\[1ex]
&\times
\left[1-(n-1)a\tau\nu^{n-1}\right]^{-n/(n-1)} .
\end{aligned}
\end{equation}
This evolution is not a simple rigid shift of the distribution toward lower frequencies, because the braking torque increases non-linearly with spin frequency. Therefore, the high-frequency tail evolves more rapidly than the low-frequency part, naturally producing a progressive narrowing of the distribution, as discussed above. Significant narrowing occurs once the following dimensionless combination is of order 1:
\begin{equation}
    (n-1)a\nu^{n-1}\tau \sim 1 .
\end{equation}

Adopting the braking strengths implied by the conservative
torque-equilibrium normalization above at
$\nu_{\rm eq}=700\,\mathrm{Hz}$, we find
that this regime of significant narrowing is reached on timescales
of order \(10^7\) yr, with only weak dependence on the specific braking mechanism. The qualitative behaviour is also only weakly sensitive to the precise choice of the equilibrium frequency. For magnetic-dipole braking, reaching this regime requires magnetic fields of order \(B\sim10^{10}\,\mathrm{G}\), larger than those typically inferred for AMXPs. In contrast, the corresponding values required for gravitational-wave-driven braking remain within astrophysically plausible ranges, namely \(\epsilon\sim10^{-8}\) for a mass quadrupole and \(\alpha_{\rm r}\sim10^{-6}\) for \(r\)-mode braking.

This suggests that deterministic gravitational-wave-driven spin-down could contribute to the formation of a narrow high-frequency population from a broader progenitor distribution. We stress, however, that this result should not be over-interpreted. The present calculation is intended to illustrate the effect of secular braking on the width of the distribution. As discussed above, a single homogeneous braking channel does not naturally reproduce the full observed distribution, which likely requires additional stochastic and/or population-dependent effects.

\section{Discussion}\label{sec:discussion}

The most obvious result of the updated analysis presented in this work is that the concentration of systems at around 570 Hz is still present. What is perhaps more interesting is that these systems are not distributed randomly in orbital period: the fastest rotators are preferentially found at longer periods. There is also some evidence that the same systems are hotter, although this result is clearly less secure. We can therefore ask whether these three observations have a common origin.

The torque argument in Sect.~\ref{sec:torques} gives us another important constraint on the interpretation. Ordinary short-timescale stochastic accretion-torque fluctuations are too weak to account for the observed widths of the spin components. Therefore, the widths of the two distributions must be population-level quantities. If the accreting pulsars are close to spin equilibrium, the observed widths must reflect a spread of effective equilibrium spin scales
\citep[see also][]{Patruno2012,Ho2014,Bhattacharyya2017,DAngelo2017}. 
If they are not close to equilibrium, the widths must instead reflect a spread of evolutionary histories, ages, accretion rates, magnetic fields, or torque states \citep[e.g.][]{2019MNRAS.488...99G,niang2026,ShaoLi2012}. 
It is important to distinguish here torque balance over an outburst from secular spin equilibrium. The former requires an instantaneous balance between accretion and braking torques during an episode of relatively high mass accretion rate. The latter requires net torques averaged over many outburst and quiescence cycles to vanish over the long-term spin evolution timescale. Therefore, measuring a small $\dot{\nu}$ during an outburst does not necessarily imply secular spin equilibrium.
There are of course observational reasons to take the equilibrium interpretation seriously. Spin equilibrium has been used successfully to describe accreting neutron-star populations over a wide range of magnetic fields and spin periods \citep{Ho2014}. Long-term timing also suggests that at least some AMXPs may be close to secular spin equilibrium
\citep[e.g.][]{Hartman2008, 2009ApJ...698L..60P, Patruno2012, Patruno2021, Sanna2022b}. This does not mean that every system in our
sample has reached torque balance, but it gives us a useful starting point.

\begin{figure*}[!th]
\centering
\includegraphics[width=\textwidth]{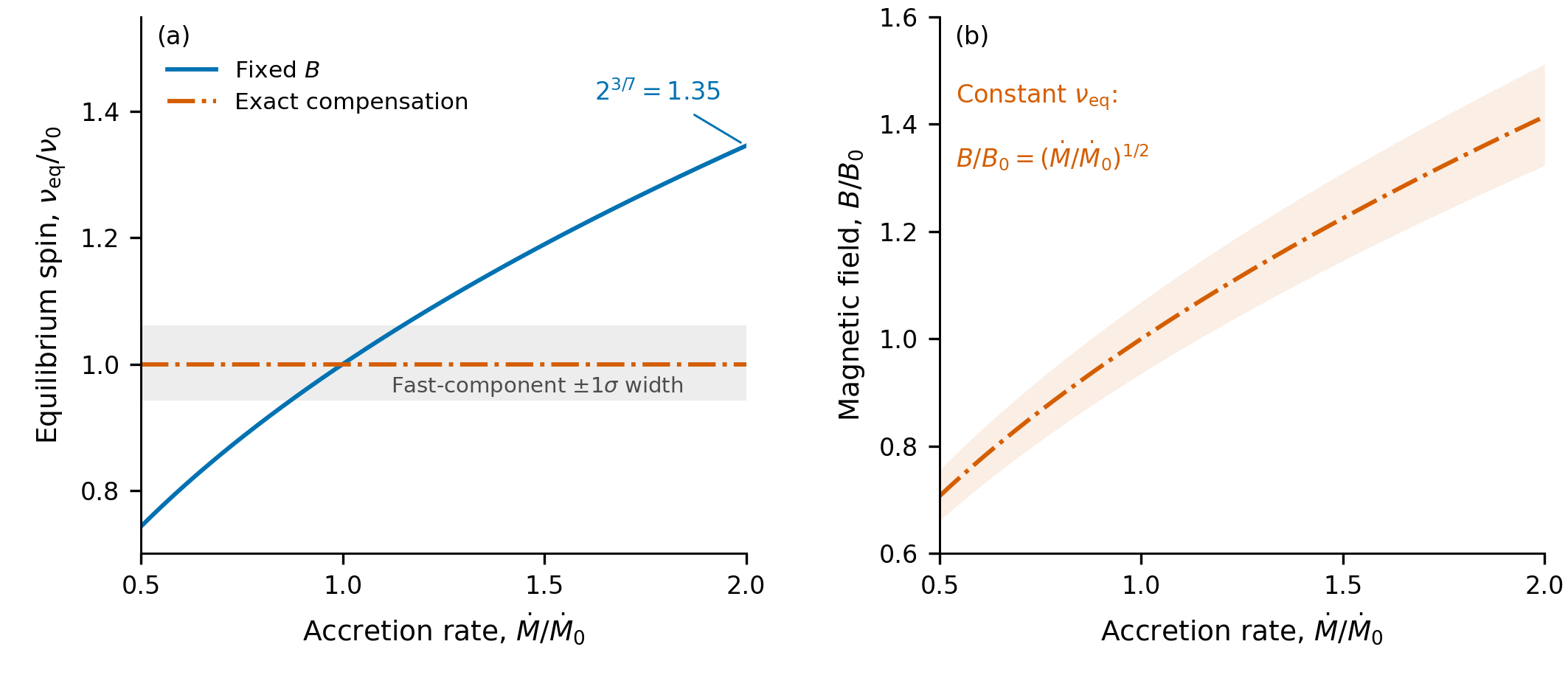}
\caption{The fast-component width and the standard equilibrium scaling $\nu_{\rm eq}\propto\dot M^{3/7}B^{-6/7}$. Reference values $\dot M_0$ and $B_0$ give $\nu_0=\nu_{\rm F}$. (a) Fixed $B$ (continuous blue) and exact compensation, ($B/B_0=(\dot M/\dot M_0)^{1/2}$) (orange dot-dashed line). Doubling $\dot M$ at fixed $B$ raises $\nu_{\rm eq}$ by a factor 1.35. The grey band shows the fitted fast-component width, $\nu_{\rm eq}/\nu_0=\exp(\pm\sigma_{\ln\nu})$, with $\sigma_{\ln\nu}=0.0572$. (b) $B$-$\dot M$ combinations within the same spin band. The orange dot-dashed line gives exactly constant $\nu_{\rm eq}$. Orange shading represents the fitted population width.
}
\label{fig:B-Mdot}
\end{figure*}
The narrow width of the fast population then becomes interesting. Using the fitted values
$\sigma_{\rm F}\simeq32.4$~Hz and $\nu_{\rm F}\simeq565.7$~Hz gives a fractional logarithmic width of
approximately
\begin{equation}
\sigma_{\ln\nu}\simeq
\frac{\sigma_{\rm F}}{\nu_{\rm F}}
\simeq 0.057 ,
\end{equation}
so the spread of the equilibrium spin frequency of the fast systems is only $\approx6$\%. If the equilibrium is set by the
magnetosphere, we can write more generally
\begin{equation}
\ln\nu_{\rm eq}
=
c+\alpha\ln\dot{M}
+\beta\ln B+\eta ,
\label{eq:general_eq_scaling}
\end{equation}
where $\eta$ represents additional source-to-source differences in the torque physics that are not explicitly included in $\dot{M}$ or $B$. Assuming for the moment that $\eta$ is uncorrelated with $\dot M$ and $B$, the variance is
\begin{equation}
\sigma_{\ln\nu}^{2}
=
\alpha^{2}\sigma_{\ln\dot{M}}^{2}
+\beta^{2}\sigma_{\ln B}^{2}
+2\alpha\beta\rho\,
 \sigma_{\ln\dot{M}}\sigma_{\ln B}
+\sigma_{\eta}^{2},
\label{eq:eq_scatter}
\end{equation}
where $\rho$ is the correlation coefficient between
$\ln\dot{M}$ and $\ln B$.
For the standard magnetospheric scaling,
$(\alpha,\beta)=(3/7,-6/7)$ \citep{GhoshLamb1979}, while
different torque prescriptions give different values. For example,
in a radiation-pressure dominated flow one has $(\alpha,\beta)=(3/14,-6/7)$ \citep[e.g.][]{2016MNRAS.457.3076D}. The simple meaning of this constraint is illustrated in Fig.~\ref{fig:B-Mdot}. In the standard case,
\begin{equation}
 \nu_{\rm eq}\propto\dot M^{3/7}B^{-6/7}.
\end{equation}
If $B$ is fixed, increasing $\dot M$ by a factor of two changes the equilibrium spin by a factor $2^{3/7}\simeq1.35$, much more than the observed width of the fast population. Increasing $B$ has the opposite effect, and the two changes cancel exactly if
$B\propto\dot M^{1/2}$. A narrow population is therefore difficult
to obtain if $\dot M$ and $B$ vary substantially and independently. Either these quantities do not vary very much among the fast systems, or their effects compensate each other to some degree.

This brings us back to the argument of~\citet{WhiteZhang1997}, who noted that the narrow spin range then inferred across LMXBs spanning a broad luminosity range implied approximately $B\propto L_X^{1/2}$ in the standard equilibrium picture if $L_X$ traced $\dot M$. \citet{Patruno2012} subsequently showed that this population-wide result was not supported by the enlarged sample of LMXBs with spin measurements, while uncertainties in disc--magnetosphere coupling further broadened the range of possible equilibrium conditions. The results presented in this paper, however, show that \textit{if the fast spin component} is composed of systems close to magnetospheric spin equilibrium as defined by Eq.~\ref{eq:general_eq_scaling}, the accreting neutron stars must have their magnetic field approximately satisfying the relation $B\propto \dot{M^{1/2}}$. 

\begin{figure*}[t]
\centering
\includegraphics[width=\textwidth]{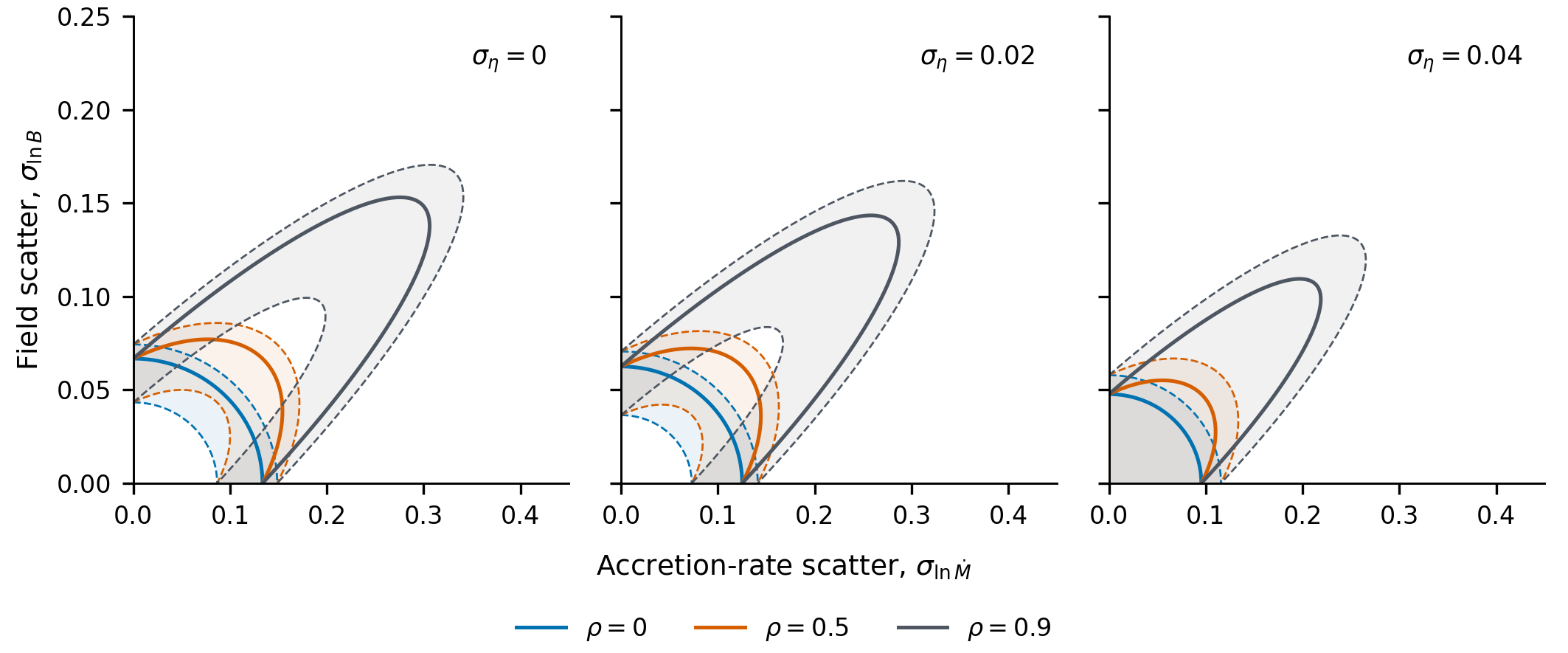}
\caption{
Population variance budget implied by the width of the fast spin component. The axes show the source-to-source logarithmic scatter in accretion rate, $\sigma_{\ln\dot M}$, and magnetic field, $\sigma_{\ln B}$, for the standard equilibrium scaling
$\nu_{\rm eq}\propto\dot M^{3/7}B^{-6/7}$. The three panels assume additional uncorrelated torque scatter $\sigma_\eta=0$, $0.02$, and $0.04$, respectively. Colours correspond to correlations $\rho=0$, $0.5$, and $0.9$ between $\ln\dot M$ and $\ln B$. Solid curves give combinations that reproduce the best-fitting fast-component width, $\sigma_{\ln\nu}=0.0572$, while the shaded bands bounded by dashed curves correspond to the 16--84\% bootstrap range of that width. Because $\dot M$ and $B$ affect the equilibrium frequency in opposite directions, a positive correlation allows larger source-to-source variations in both quantities while preserving a narrow spin distribution. Additional uncorrelated torque scatter instead consumes part of the observed variance budget and reduces the allowed scatter in $\dot M$ and $B$. The contours are conditional variance relations, not measurements or confidence regions for $\dot M$, $B$, or their correlation.
}
\label{fig:variance-budget}
\end{figure*}

However, a complication is introduced by the effect of covariance and correlation between components as described in Eq.~\ref{eq:eq_scatter}. 
Since $\alpha\beta<0$ in the standard scaling, a positive
correlation between $\dot M$ and $B$ reduces the resulting spin variance, whereas independent variations add in quadrature. Any additional uncorrelated source-to-source torque scatter, represented by $\sigma_\eta$, consumes part of the small observed variance budget and therefore leaves less room for
variation in $\dot M$ and $B$. Figure~\ref{fig:variance-budget} quantifies the combinations of scatter and correlation compatible with the observed fast component. For $\rho=0$, the narrow spin width permits only a relatively small scatter in either quantity, whereas increasing $\rho$ allows a broader range of both $\dot M$ and $B$ because their contributions increasingly cancel out. Increasing $\sigma_\eta$ has the opposite effect and contracts the range of
$\sigma_{\ln\dot M}$ and $\sigma_{\ln B}$ compatible with the observed fast branch. What matters, therefore, is not the spread in $\dot M$ or $B$ separately, but the spread in their combination that sets $\nu_{\rm eq}$.

This is also relevant for continuous GW searches. Torque-balance arguments are often used as a benchmark for the GW strain that may be expected from these systems, simply estimating the GW spin-down torque that would be required to offset the spin-up torque due to accretion \citep{Watts2008GW, RilesReview}. Our results suggest that such arguments may be more relevant for
the fast population, where the narrow spin accumulation points to a regulated torque state, than for the full LMXB population.

We can look at the same problem in a different way with the calculation of Sect.~\ref{sec:braking}. Starting from a Gaussian spin distribution, we can ask how its shape changes when we add a coherent braking torque and stochastic fluctuations.
Deterministic braking shifts systems toward lower spin and can sharpen the high-frequency edge, while stochastic fluctuations broaden and flatten the distribution. However, the slow-branch-like flat morphologies require stochastic broadening much stronger than ordinary accretion-torque noise, by several orders of magnitude, so that this points either against this mechanism or requires an unknown very strong source of stochastic variation. Narrow and peaked morphologies occur instead when deterministic braking dominates. Thus, a single homogeneous stochastic braking channel does not naturally produce both the broad slow branch and the narrow fast branch as long-lived
features. This does not exclude gravitational-wave torques, because coherent gravitational-wave braking can still act as a high-frequency regulator and may help explain the
absence of systems above $\simeq600$--$700$ Hz. Previous population-synthesis work \citep{2019MNRAS.488...99G} has shown that a relatively well-defined gravitational-wave torque
amplitude can produce a high-frequency pile-up. This is consistent with the constraint derived here: the fast branch can remain narrow only if the effective regulating torque has limited source-to-source scatter. The point is that a high-frequency cutoff alone is not enough. Any explanation for the spin distribution should also account for the orbital-period difference and, possibly, the temperature difference.

The orbital periods show an interesting pattern. All the short-period systems are slow rotators, while at longer periods we find both slow and fast systems. A long orbital period is therefore not sufficient to produce a fast rotator. There must be something else that distinguishes these systems. The obvious possibilities are the secular mass-transfer rate, the magnetic field, outflows, or perhaps an additional braking torque. At present we cannot distinguish between these alternatives. 

The temperature difference may provide another clue, although the present data are not sufficient to identify the mechanism. In the deep-crustal-heating picture \citep[e.g.][]{Yakovlev2004_rev}, the time-averaged quiescent luminosity roughly scales as:
\begin{equation}
L_q \simeq \langle\dot M\rangle Q_{\rm nuc},
\end{equation}
with $Q_{\rm nuc} \sim 1$--$2~\mathrm{MeV}$ per accreted nucleon, while the interior temperature is set by the competition between this heating and neutrino cooling in the core. In this scenario, the hotter fast branch would point to a larger long-term average accretion rate and therefore stronger deep-crustal heating. However, this is not the only interpretation capable of explaining the observations. Different neutrino-cooling efficiencies could also contribute to the observed temperature difference~\citep[e.g.][]{Yakovlev2004,Potekhin2019,Brown2018,Mendes2022}. Furthermore, differences in the NS envelope composition can modify the mapping between core temperature and observed surface temperature
\citep[e.g.][]{Potekhin1997,Brown2002,Yakovlev2004,cavecchi2026,nava-callejas2026}. 

For $r$-mode models the thermal information is especially relevant. An $r$-mode capable of providing a substantial spin-down torque also dissipates energy inside the star. An $r$-mode interpretation of the fast population must therefore be consistent simultaneously with the required torque and with the thermal balance of the neutron star. The present surface-temperature comparison is not sufficiently precise to perform that test by itself, but it provides an independent constraint that any such model needs to satisfy.

Finally, low-level residual accretion during quiescence can produce variable thermal and non-thermal emission, which may contaminate a simple interpretation of quiescent temperature as a pure cooling observable \citep[e.g.][]{Cackett2010,Cackett2011,Cackett2012,thermalParikh2017}, and could potentially point to some spin-regulating event occurring during quiescence. 
We therefore have evidence that the spin evolution is connected to the long-term evolution of the binary. We cannot identify the mechanism with the present data, but whatever sets the spin of the fast systems must operate over a much narrower range of conditions than in the slow population.

\section{Conclusions}

We have revisited the spin distribution of 39 accreting neutron stars in
LMXBs, extending the sample analysed by \citet{2017ApJ...850..106P}. The
preferred empirical description contains a broad population at lower spin frequencies together with a much narrower concentration centred near $\simeq 566$~Hz, with a width of about 32~Hz. The existence and location of this high-frequency concentration are robust to the choice of mixture model, whereas the precise division between the fast and slow populations is much less well determined. With the enlarged sample we also find that spin is related to other properties of the binaries. Fast rotators preferentially occur at longer orbital periods, and the same ordering is present when only AMXPs are considered. The quiescent-temperature comparison provides additional, but less secure, evidence that the fast systems tend to be hotter. 

The different widths of the two components also tell us something about their spin evolution. Ordinary short-timescale accretion-torque fluctuations are too weak to generate the broad slow component and cannot account for most of the best-fitting width of the fast component. The observed widths must therefore mainly reflect source-to-source differences in long-term
torque conditions and/or different evolutionary histories. If the fast population is close to spin equilibrium, its narrow width implies that the combination of accretion rate, magnetic field, and any additional braking torque that determines the equilibrium spin must vary relatively little across these systems.

Coherent nonlinear braking can also compress an initially broader
high-frequency population, but the spin distribution alone does not identify the braking mechanism. Magnetospheric regulation, correlated variations in accretion rate and magnetic field, and additional spin-down torques, including gravitational-wave emission, all remain possible. 

Any successful evolutionary model of accreting millisecond neutron stars should reproduce these properties together. The data suggest that the long-term torque history of the neutron star is connected to the evolution and accretion history of the binary. With the present sample, however, we cannot determine which physical mechanism produces this connection.

\begin{acknowledgements}
AP acknowledges support from the Spanish Ministerio de Ciencia e Innovación under grant PID2024-155316NB-I00. Y.C. acknowledges support from the grant RYC2021-032718-I, funded by MCIN/AEI/10.13039/501100011033 and the European Union NextGenerationEU/PRTR. 
M.F-M. was supported by the yearly plan of the Tourist Stay Tax for 2023 (ITS2023-086 - Programa de Fomento de la Investigación), the Universitat de les Illes Balears (UIB) with funds from the Programa de Foment de la Recerca i la Innovació de la UIB 2024-2026 (supported by the yearly plan of the Tourist Stay Tax ITS2023-086); the Spanish Agencia Estatal de Investigación grants PID2022-138626NB-I00, RED2024-153978-E, RED2024-153735-E, funded by MICIU/AEI/10.13039/501100011033 and the ERDF/EU; and the Comunitat Autònoma de les Illes Balears through the Conselleria d'Educació i Universitats with funds from the ERDF (SINCO2022/18146 - Plataforma HiTech-IAC3-BIO). This document has been assigned LIGO document number \href{https://dcc.ligo.org/P2600519}{P2600519}.

\end{acknowledgements}

\bibliographystyle{aa}
\bibliography{sample631_final}

@ARTICLE{2016MNRAS.457.3076D,
       author = {{Dall'Osso}, Simone and {Perna}, Rosalba and {Papitto}, Alessandro and {Bozzo}, Enrico and {Stella}, Luigi},
        title = "{The accretion regimes of a highly magnetized NS: the unique case of NuSTAR J095551+6940.8}",
      journal = {\mnras},
         year = 2016,
        month = apr,
       volume = {457},
       number = {3},
        pages = {3076-3083},
          doi = {10.1093/mnras/stw110},
archivePrefix = {arXiv},
       eprint = {1512.01532},
 primaryClass = {astro-ph.HE},
       adsurl = {https://ui.adsabs.harvard.edu/abs/2016MNRAS.457.3076D}
}

@INPROCEEDINGS{Papitto2022,
       author = {{Papitto}, Alessandro and {de Martino}, Domitilla},
        title = "{Transitional Millisecond Pulsars}",
    booktitle = {Astrophysics and Space Science Library},
         year = 2022,
       editor = {{Bhattacharyya}, Sudip and {Papitto}, Alessandro and {Bhattacharya}, Dipankar},
       series = {Astrophysics and Space Science Library},
       volume = {465},
        month = jan,
        pages = {157-200},
          doi = {10.1007/978-3-030-85198-9_6},
archivePrefix = {arXiv},
       eprint = {2010.09060},
 primaryClass = {astro-ph.HE},
       adsurl = {https://ui.adsabs.harvard.edu/abs/2022ASSL..465..157P}
}

@article{2017ApJ...850..106P,
  author = {{Patruno}, A. and {Haskell}, B. and {Andersson}, N.},
  title = {{The Spin Distribution of Fast-spinning Neutron Stars in Low-mass X-Ray Binaries: Evidence for Two Subpopulations}},
  journal = {\apj},
  year = {2017},
  month = {nov},
  volume = {850},
  number = {1},
  pages = {106},
  doi = {10.3847/1538-4357/aa927a},
  eprint = {1705.07669},
  archivePrefix = {arXiv},
  primaryClass = {astro-ph.HE}
}

@article{2019MNRAS.488...99G,
  author = {{Gittins}, Fabian and {Andersson}, Nils},
  title = {{Population synthesis of accreting neutron stars emitting gravitational waves}},
  journal = {\mnras},
  year = {2019},
  month = {sep},
  volume = {488},
  number = {1},
  pages = {99--110},
  doi = {10.1093/mnras/stz1719},
  eprint = {1811.00550},
  archivePrefix = {arXiv},
  primaryClass = {astro-ph.HE}
}

@article{2022ApJ...941L..30A,
  author = {{Abbott}, R. and {Abe}, H. and {Acernese}, F. and others},
  title = {{Model-based Cross-correlation Search for Gravitational Waves from the Low-mass X-Ray Binary Scorpius X-1 in LIGO O3 Data}},
  journal = {\apjl},
  year = {2022{\natexlab{b}}},
  month = {dec},
  volume = {941},
  number = {2},
  pages = {L30},
  doi = {10.3847/2041-8213/aca1b0},
  eprint = {2209.02863},
  archivePrefix = {arXiv},
  primaryClass = {astro-ph.HE}
}

@article{Alpar1982,
  author = {{Alpar}, M.~A. and {Cheng}, A.~F. and {Ruderman}, M.~A. and {Shaham}, J.},
  title = {{A new class of radio pulsars}},
  journal = {\nat},
  year = {1982},
  month = {dec},
  volume = {300},
  number = {5894},
  pages = {728--730},
  doi = {10.1038/300728a0}
}

@article{Andersson1998,
  author = {Andersson, N.},
  title = {A new class of unstable modes of rotating relativistic stars},
  journal = {The Astrophysical Journal},
  year = {1998},
  volume = {502},
  pages = {708--713},
  doi = {10.1086/305919}
}

@article{AnderssonKokkotasSchutz1999,
  author = {Andersson, N. and Kokkotas, K. D. and Schutz, B. F.},
  title = {Gravitational radiation limit on the spin of young neutron stars},
  journal = {The Astrophysical Journal},
  year = {1999},
  volume = {510},
  pages = {846--853},
  doi = {10.1086/306625}
}

@article{Archibald2009_J1023,
  author = {{Archibald}, Anne M. and {Stairs}, Ingrid H. and {Ransom}, Scott M. and {Kaspi}, Victoria M. and {Kondratiev}, Vladislav I. and {Lorimer}, Duncan R. and {McLaughlin}, Maura A. and {Boyles}, Jason and {Hessels}, Jason W. T. and {Lynch}, Ryan and {van Leeuwen}, Joeri and {Roberts}, Mallory S. E. and {Jenet}, Frederick and {Champion}, David J. and {Rosen}, Rachel and {Barlow}, Brad N. and {Dunlap}, Bart H. and {Remillard}, Ronald A.},
  title = {A Radio Pulsar/X-ray Binary Link},
  journal = {Science},
  year = {2009},
  volume = {324},
  pages = {1411--1414},
  doi = {10.1126/science.1172740},
  eprint = {0905.3397},
  archivePrefix = {arXiv}
}

@article{ArmasPadilla2013_17494,
  author = {{Armas Padilla}, M. and Wijnands, R. and Degenaar, N.},
  title = {{XMM--Newton and Swift spectroscopy of the newly discovered very faint X-ray transient IGR J17494--3030}},
  journal = {Monthly Notices of the Royal Astronomical Society: Letters},
  year = {2013},
  volume = {436},
  pages = {L89--L93},
  doi = {10.1093/mnrasl/slt119},
  url = {https://academic.oup.com/mnrasl/article/436/1/L89/989968}
}

@article{Beri2023_XTEJ1739,
  author = {Beri, Aru and Sharma, Rahul and Roy, Pinaki and Gaur, Vishal and Altamirano, Diego and Andersson, Nils and Gittins, Fabian and Celora, T.},
  title = {AstroSat and NuSTAR observations of XTE J1739-285 during the 2019--2020 outburst},
  journal = {Monthly Notices of the Royal Astronomical Society},
  year = {2023},
  volume = {521},
  pages = {5904--5916},
  doi = {10.1093/mnras/stad902},
  eprint = {2303.13085},
  archivePrefix = {arXiv}
}

@article{BhattacharyaVanDenHeuvel1991,
  author = {Bhattacharya, D. and van den Heuvel, E. P. J.},
  title = {Formation and evolution of binary and millisecond radio pulsars},
  journal = {Physics Reports},
  year = {1991},
  volume = {203},
  pages = {1--124},
  doi = {10.1016/0370-1573(91)90064-S}
}

@article{Bhattacharyya2017,
  author = {{Bhattacharyya}, Sudip and {Chakrabarty}, Deepto},
  title = {{The Effect of Transient Accretion on the Spin-up of Millisecond Pulsars}},
  journal = {\apj},
  year = {2017},
  month = {jan},
  volume = {835},
  number = {1},
  pages = {4},
  doi = {10.3847/1538-4357/835/1/4},
  eprint = {1612.04962},
  archivePrefix = {arXiv},
  primaryClass = {astro-ph.HE}
}

@article{Bildsten1998,
  author = {{Bildsten}, Lars},
  title = {{Gravitational Radiation and Rotation of Accreting Neutron Stars}},
  journal = {\apjl},
  year = {1998},
  month = {jul},
  volume = {501},
  number = {1},
  pages = {L89--L93},
  doi = {10.1086/311440},
  eprint = {astro-ph/9804325},
  archivePrefix = {arXiv},
  primaryClass = {astro-ph}
}

@article{Brown1998,
  author = {{Brown}, Edward F. and {Bildsten}, Lars and {Rutledge}, Robert E.},
  title = {{Crustal Heating and Quiescent Emission from Transiently Accreting Neutron Stars}},
  journal = {\apjl},
  year = {1998},
  month = {sep},
  volume = {504},
  number = {2},
  pages = {L95--L98},
  doi = {10.1086/311578},
  eprint = {astro-ph/9807179},
  archivePrefix = {arXiv},
  primaryClass = {astro-ph}
}

@article{Brown2002,
  author = {Brown, Edward F. and Bildsten, Lars and Chang, Philip},
  title = {Variability in the Thermal Emission from Accreting Neutron Star Transients},
  journal = {The Astrophysical Journal},
  year = {2002},
  volume = {574},
  pages = {920--929},
  doi = {10.1086/341066},
  eprint = {astro-ph/0204102},
  archivePrefix = {arXiv}
}

@article{Brown2018,
  author = {Brown, Edward F. and Cumming, Andrew and Fattoyev, Farrukh J. and Horowitz, C. J. and Page, Dany and Reddy, Sanjay},
  title = {Rapid neutrino cooling in the neutron star MXB 1659-29},
  journal = {Physical Review Letters},
  year = {2018},
  volume = {120},
  pages = {182701},
  doi = {10.1103/physrevlett.120.182701},
  eprint = {1801.00041},
  archivePrefix = {arXiv},
  primaryClass = {astro-ph.HE}
}

@article{Bult2019_IGRJ17379,
  author = {{Bult}, Peter and {Markwardt}, Craig B. and {Altamirano}, Diego and {Arzoumanian}, Zaven and {Chakrabarty}, Deepto and {Gendreau}, Keith C. and {Guillot}, Sebastien and {Jaisawal}, Gaurava K. and {Ray}, Paul. S. and {Strohmayer}, Tod E.},
  title = {On the Curious Pulsation Properties of the Accreting Millisecond Pulsar IGR J17379-3747},
  journal = {The Astrophysical Journal},
  year = {2019},
  volume = {877},
  pages = {70},
  doi = {10.3847/1538-4357/ab1b26},
  eprint = {1904.11311},
  archivePrefix = {arXiv}
}

@article{Bult2021_IGRJ17062,
  author = {Bult, Peter and Strohmayer, Tod E. and Malacaria, Christian and Ng, Mason and Wadiasingh, Zorawar},
  title = {Long-term coherent timing of the accreting millisecond pulsar IGR J17062-6143},
  journal = {The Astrophysical Journal},
  year = {2021},
  volume = {912},
  pages = {120},
  doi = {10.3847/1538-4357/abf13f},
  eprint = {2103.12556},
  archivePrefix = {arXiv}
}

@article{Bult2022_MAXIJ1816,
  author = {{Bult}, Peter and {Altamirano}, Diego and {Arzoumanian}, Zaven and {Chakrabarty}, Deepto and {Chenevez}, J{\'{e}}r{\^{o}}me and {Ferrara}, Elizabeth C. and {Gendreau}, Keith C. and {Guillot}, Sebastien and {G{\"{u}}ver}, Tolga and {Iwakiri}, Wataru and {Jaisawal}, Gaurava K. and {Mancuso}, Giulio C. and {Malacaria}, Christian and {Ng}, Mason and {Sanna}, Andrea and {Strohmayer}, Tod E. and {Wadiasingh}, Zorawar and {Wolff}, Michael T.},
  title = {The Discovery of the 528.6 Hz Accreting Millisecond X-Ray Pulsar MAXI J1816-195},
  journal = {The Astrophysical Journal Letters},
  year = {2022},
  volume = {935},
  pages = {L32},
  doi = {10.3847/2041-8213/ac87f9},
  eprint = {2208.04721},
  archivePrefix = {arXiv}
}

@article{Cackett2010,
  author = {Cackett, Edward M. and Brown, Edward F. and Miller, Jon M. and Wijnands, Rudy},
  title = {Quiescent X-ray emission from Cen X-4: a variable thermal component},
  journal = {The Astrophysical Journal},
  year = {2010},
  volume = {720},
  pages = {1325--1332},
  doi = {10.1088/0004-637X/720/2/1325},
  eprint = {1007.2823},
  archivePrefix = {arXiv},
  primaryClass = {astro-ph.HE}
}

@article{Cackett2011,
  author = {Cackett, E. M. and Fridriksson, J. K. and Homan, J. and Miller, J. M. and Wijnands, R.},
  title = {Quiescent X-ray variability from the neutron star transient Aql X-1},
  journal = {Monthly Notices of the Royal Astronomical Society},
  year = {2011},
  volume = {414},
  pages = {3006--3013},
  doi = {10.1111/j.1365-2966.2011.18601.x},
  eprint = {1102.5016},
  archivePrefix = {arXiv},
  primaryClass = {astro-ph.HE}
}

@article{Cackett2012,
  author = {Cackett, E. M. and Brown, E. F. and Degenaar, N. and Miller, J. M. and Reynolds, M. and Wijnands, R.},
  title = {An X-ray--UV correlation in Cen X-4 during quiescence},
  journal = {Monthly Notices of the Royal Astronomical Society},
  year = {2013},
  volume = {433},
  pages = {1362--1368},
  doi = {10.1093/mnras/stt804},
  eprint = {1210.4510},
  archivePrefix = {arXiv},
  primaryClass = {astro-ph.HE}
}

@article{Chakrabarty2003,
  author = {{Chakrabarty}, Deepto and {Morgan}, Edward H. and {Muno}, Michael P. and {Galloway}, Duncan K. and {Wijnands}, Rudy and {van der Klis}, Michiel and {Markwardt}, Craig B.},
  title = {{Nuclear-powered millisecond pulsars and the maximum spin frequency of neutron stars}},
  journal = {\nat},
  year = {2003},
  month = {jul},
  volume = {424},
  number = {6944},
  pages = {42--44},
  doi = {10.1038/nature01732},
  eprint = {astro-ph/0307029},
  archivePrefix = {arXiv},
  primaryClass = {astro-ph}
}

@article{Chakrabarty2013_4886,
  author = {Chakrabarty, D. and Jonker, P. G. and Markwardt, C. B.},
  title = {{2012 Chandra Non-Detection of the Fading X-ray Transient IGR J17494--3030}},
  journal = {The Astronomer's Telegram},
  year = {2013},
  volume = {4886},
  pages = {1},
  url = {https://www.astronomerstelegram.org/?read=4886}
}

@article{Cook1994,
  author = {{Cook}, Gregory B. and {Shapiro}, Stuart L. and {Teukolsky}, Saul A.},
  title = {{Rapidly Rotating Neutron Stars in General Relativity: Realistic Equations of State}},
  journal = {\apj},
  year = {1994},
  month = {apr},
  volume = {424},
  pages = {823},
  doi = {10.1086/173934}
}

@article{DAngelo2017,
  author = {{D'Angelo}, Caroline R.},
  title = {Spin equilibrium in strongly magnetized accreting stars},
  journal = {Monthly Notices of the Royal Astronomical Society},
  year = {2017},
  volume = {470},
  number = {3},
  pages = {3316--3331},
  doi = {10.1093/mnras/stx1306},
  eprint = {1609.08654},
  archivePrefix = {arXiv},
  primaryClass = {astro-ph.HE}
}

@article{DAngeloSpruit2010,
  author = {D'Angelo, C. R. and Spruit, H. C.},
  title = {Episodic accretion on to strongly magnetic stars},
  journal = {Monthly Notices of the Royal Astronomical Society},
  year = {2010},
  volume = {406},
  pages = {1208--1219},
  doi = {10.1111/j.1365-2966.2010.16749.x},
  eprint = {1001.1742},
  archivePrefix = {arXiv}
}

@article{DaiLi2006,
  author = {Dai, H.-L. and Li, X.-D.},
  title = {Accretion torque on magnetized neutron stars},
  journal = {Astronomy and Astrophysics},
  year = {2006},
  volume = {451},
  pages = {581--585},
  doi = {10.1051/0004-6361:20053907}
}

@article{Dangelo2012,
  author = {{D'Angelo}, Caroline R. and {Spruit}, Hendrik C.},
  title = {{Accretion discs trapped near corotation}},
  journal = {\mnras},
  year = {2012},
  month = {feb},
  volume = {420},
  number = {1},
  pages = {416--429},
  doi = {10.1111/j.1365-2966.2011.20046.x},
  eprint = {1108.3833},
  archivePrefix = {arXiv},
  primaryClass = {astro-ph.SR}
}

@article{Deloye2003,
  author = {{Deloye}, Christopher J. and {Bildsten}, Lars},
  title = {{White Dwarf Donors in Ultracompact Binaries: The Stellar Structure of Finite-Entropy Objects}},
  journal = {\apj},
  year = {2003},
  month = {dec},
  volume = {598},
  number = {2},
  pages = {1217--1228},
  doi = {10.1086/379063},
  eprint = {astro-ph/0308233},
  archivePrefix = {arXiv},
  primaryClass = {astro-ph}
}

@article{Echeveste2024,
  author = {{Echeveste}, M. and {Novarino}, M.~L. and {Benvenuto}, O.~G. and {De Vito}, M.~A.},
  title = {{The impact of different magnetic braking prescriptions on the evolution of LMXBs}},
  journal = {\mnras},
  year = {2024},
  month = {jun},
  volume = {530},
  number = {4},
  pages = {4277--4287},
  doi = {10.1093/mnras/stae1115},
  eprint = {2404.16185},
  archivePrefix = {arXiv},
  primaryClass = {astro-ph.HE}
}

@article{GhoshLamb1979,
  author = {{Ghosh}, P. and {Lamb}, F.~K.},
  title = {{Accretion by rotating magnetic neutron stars. III. Accretion torques and period changes in pulsating X-ray sources.}},
  journal = {\apj},
  year = {1979},
  month = {nov},
  volume = {234},
  pages = {296-316},
  doi = {10.1086/157498}
}

@article{Gusinskaia2020_17591,
  author = {Gusinskaia, N. V. and Russell, T. D. and Hessels, J. W. T. and others},
  title = {{Radio and X-ray monitoring of the accreting millisecond X-ray pulsar IGR J17591--2342 in outburst}},
  journal = {Monthly Notices of the Royal Astronomical Society},
  year = {2020},
  volume = {492},
  pages = {1091--1101},
  doi = {10.1093/mnras/stz3460},
  eprint = {1909.02323},
  archivePrefix = {arXiv},
  url = {https://arxiv.org/abs/1909.02323}
}

@article{Hartman2008,
  author = {Hartman, J. M. and Patruno, A. and Chakrabarty, D. and Kaplan, D. L. and Markwardt, C. B. and Morgan, E. H. and Ray, P. S. and van der Klis, M. and Wijnands, R.},
  title = {The long-term evolution of the spin, pulse shape, and orbit of the accretion-powered millisecond pulsar SAX J1808.4-3658},
  journal = {The Astrophysical Journal},
  year = {2008},
  volume = {675},
  pages = {1468--1486},
  doi = {10.1086/527461},
  eprint = {0708.0211},
  archivePrefix = {arXiv}
}

@article{Haskell2008,
  author = {{Haskell}, B. and {Samuelsson}, L. and {Glampedakis}, K. and {Andersson}, N.},
  title = {{Modelling magnetically deformed neutron stars}},
  journal = {\mnras},
  year = {2008},
  month = {mar},
  volume = {385},
  number = {1},
  pages = {531--542},
  doi = {10.1111/j.1365-2966.2008.12861.x},
  eprint = {0705.1780},
  archivePrefix = {arXiv},
  primaryClass = {astro-ph}
}

@article{HaskellRev,
  author = {{Haskell}, B. and {Bejger}, M.},
  title = {{Astrophysics with continuous gravitational waves}},
  journal = {Nature Astronomy},
  year = {2023},
  month = {oct},
  volume = {7},
  pages = {1160--1170},
  doi = {10.1038/s41550-023-02059-w}
}

@article{HaskellZdunik18,
  author = {{Haskell}, B. and {Zdunik}, J.~L. and {Fortin}, M. and {Bejger}, M. and {Wijnands}, R. and {Patruno}, A.},
  title = {{Fundamental physics and the absence of sub-millisecond pulsars}},
  journal = {\aap},
  year = {2018},
  month = {dec},
  volume = {620},
  pages = {A69},
  doi = {10.1051/0004-6361/201833521},
  eprint = {1805.11277},
  archivePrefix = {arXiv},
  primaryClass = {astro-ph.HE}
}

@article{Ho2014,
  author = {Ho, W. C. G. and Klus, H. and Coe, M. J. and Andersson, N.},
  title = {Equilibrium spin pulsars unite neutron star populations},
  journal = {Monthly Notices of the Royal Astronomical Society},
  year = {2014},
  volume = {437},
  pages = {3664--3669},
  doi = {10.1093/mnras/stt2193},
  eprint = {1311.1969},
  archivePrefix = {arXiv},
  primaryClass = {astro-ph.HE}
}

@article{Knigge2011,
  author = {{Knigge}, Christian and {Baraffe}, Isabelle and {Patterson}, Joseph},
  title = {{The Evolution of Cataclysmic Variables as Revealed by Their Donor Stars}},
  journal = {\apjs},
  year = {2011},
  month = {jun},
  volume = {194},
  number = {2},
  pages = {28},
  doi = {10.1088/0067-0049/194/2/28},
  eprint = {1102.2440},
  archivePrefix = {arXiv},
  primaryClass = {astro-ph.SR}
}

@article{LattimerPrakash2007,
  author = {Lattimer, J. M. and Prakash, M.},
  title = {Neutron star observations: Prognosis for equation of state constraints},
  journal = {Physics Reports},
  year = {2007},
  volume = {442},
  pages = {109--165},
  doi = {10.1016/j.physrep.2007.02.003}
}

@article{MelatosPayne2005,
  author = {{Melatos}, A. and {Payne}, D.~J.~B.},
  title = {{Gravitational Radiation from an Accreting Millisecond Pulsar with a Magnetically Confined Mountain}},
  journal = {\apj},
  year = {2005},
  month = {apr},
  volume = {623},
  number = {2},
  pages = {1044--1050},
  doi = {10.1086/428600},
  eprint = {astro-ph/0503287},
  archivePrefix = {arXiv},
  primaryClass = {astro-ph}
}

@article{Mendes2022,
  author = {{Mendes}, Melissa and {Fattoyev}, Farrukh J. and {Cumming}, Andrew and {Gale}, Charles},
  title = {{Fast Neutrino Cooling in the Accreting Neutron Star MXB 1659-29}},
  journal = {\apj},
  year = {2022},
  month = {oct},
  volume = {938},
  number = {2},
  pages = {119},
  doi = {10.3847/1538-4357/ac9138},
  eprint = {2208.04262},
  archivePrefix = {arXiv},
  primaryClass = {astro-ph.HE}
}

@article{Molkov2024_SRGAJ144459,
  author = {{Molkov}, Sergey V. and {Lutovinov}, Alexander A. and {Tsygankov}, Sergey S. and {Suleimanov}, Valery F. and {Poutanen}, Juri and {Lapshov}, Igor Yu. and {Mereminskiy}, Ilya A. and {Semena}, Andrei N. and {Arefiev}, Vadim A. and {Tkachenko}, Alexey Yu.},
  title = {{Discovery of SRGA J144459.2{\ensuremath{-}}604207 with the SRG/ART-XC telescope: A well-tempered bursting accreting millisecond X-ray pulsar}},
  journal = {\aap},
  year = {2024},
  month = {oct},
  volume = {690},
  pages = {A353},
  doi = {10.1051/0004-6361/202450581},
  eprint = {2404.19709},
  archivePrefix = {arXiv},
  primaryClass = {astro-ph.HE}
}

@article{NelemansJonker2010,
  author = {{Nelemans}, G. and {Jonker}, P.~G.},
  title = {{Ultra-compact (X-ray) binaries}},
  journal = {\nar},
  year = {2010},
  month = {mar},
  volume = {54},
  number = {3-6},
  pages = {87--92},
  doi = {10.1016/j.newar.2010.09.021},
  eprint = {astro-ph/0605722},
  archivePrefix = {arXiv},
  primaryClass = {astro-ph}
}

@article{Ng2021_IGRJ17494,
  author = {{Ng}, Mason and {Ray}, Paul S. and {Bult}, Peter and {Chakrabarty}, Deepto and {Jaisawal}, Gaurava K. and {Malacaria}, Christian and {Altamirano}, Diego and {Arzoumanian}, Zaven and {Gendreau}, Keith C. and {G{\"{u}}ver}, Tolga and {Kerr}, Matthew and {Strohmayer}, Tod E. and {Wadiasingh}, Zorawar and {Wolff}, Michael T.},
  title = {NICER Discovery of Millisecond X-Ray Pulsations and an Ultracompact Orbit in IGR J17494-3030},
  journal = {The Astrophysical Journal Letters},
  year = {2021},
  volume = {908},
  pages = {L15},
  doi = {10.3847/2041-8213/abe1b4},
  eprint = {2101.06324},
  archivePrefix = {arXiv}
}

@article{Ng2024_SRGA_thermal,
  author = {Ng, M. and Ray, P. S. and Sanna, A. and others},
  title = {{NICER Discovery that SRGA J144459.2--604207 Is an Accreting Millisecond X-Ray Pulsar}},
  journal = {The Astrophysical Journal Letters},
  year = {2024},
  volume = {968},
  pages = {L7},
  doi = {10.3847/2041-8213/ad4edb},
  eprint = {2405.00087},
  archivePrefix = {arXiv},
  url = {https://arxiv.org/abs/2405.00087}
}

@article{Owen1998,
  author = {{Owen}, Benjamin J. and {Lindblom}, Lee and {Cutler}, Curt and {Schutz}, Bernard F. and {Vecchio}, Alberto and {Andersson}, Nils},
  title = {{Gravitational waves from hot young rapidly rotating neutron stars}},
  journal = {\prd},
  year = {1998},
  month = {oct},
  volume = {58},
  number = {8},
  pages = {084020},
  doi = {10.1103/PhysRevD.58.084020},
  eprint = {gr-qc/9804044},
  archivePrefix = {arXiv},
  primaryClass = {gr-qc}
}

@article{Ozel2009,
  author = {{{\"O}zel}, Feryal and {Psaltis}, Dimitrios},
  title = {{Reconstructing the neutron-star equation of state from astrophysical measurements}},
  journal = {\prd},
  year = {2009},
  month = {nov},
  volume = {80},
  number = {10},
  pages = {103003},
  doi = {10.1103/PhysRevD.80.103003},
  eprint = {0905.1959},
  archivePrefix = {arXiv},
  primaryClass = {astro-ph.HE}
}

@article{Papitto2014,
  author = {{Papitto}, A. and {Torres}, D.~F. and {Rea}, N. and {Tauris}, T.~M.},
  title = {{Spin frequency distributions of binary millisecond pulsars}},
  journal = {\aap},
  year = {2014},
  month = {jun},
  volume = {566},
  pages = {A64},
  doi = {10.1051/0004-6361/201321724},
  eprint = {1403.6775},
  archivePrefix = {arXiv},
  primaryClass = {astro-ph.HE}
}

@article{Patruno2010,
  author = {{Patruno}, Alessandro},
  title = {{The Accreting Millisecond X-ray Pulsar IGR J00291+5934: Evidence for a Long Timescale Spin Evolution}},
  journal = {\apj},
  year = {2010},
  month = {oct},
  volume = {722},
  number = {1},
  pages = {909--918},
  doi = {10.1088/0004-637X/722/1/909},
  eprint = {1006.0815},
  archivePrefix = {arXiv},
  primaryClass = {astro-ph.HE}
}

@article{Patruno2012,
  author = {{Patruno}, Alessandro and {Haskell}, Brynmor and {D'Angelo}, Caroline},
  title = {{Gravitational Waves and the Maximum Spin Frequency of Neutron Stars}},
  journal = {\apj},
  year = {2012},
  month = {feb},
  volume = {746},
  number = {1},
  pages = {9},
  doi = {10.1088/0004-637X/746/1/9},
  eprint = {1109.0536},
  archivePrefix = {arXiv},
  primaryClass = {astro-ph.HE}
}

@article{Patruno2016,
  author = {{Patruno}, A. and {Maitra}, D. and {Curran}, P.~A. and {D'Angelo}, C. and {Fridriksson}, J.~K. and {Russell}, D.~M. and {Middleton}, M. and {Wijnands}, R.},
  title = {{The Reflares and Outburst Evolution in the Accreting Millisecond Pulsar SAX J1808.4-3658: A Disk Truncated Near Co-Rotation?}},
  journal = {\apj},
  year = {2016},
  month = {feb},
  volume = {817},
  number = {2},
  pages = {100},
  doi = {10.3847/0004-637X/817/2/100},
  eprint = {1504.05048},
  archivePrefix = {arXiv},
  primaryClass = {astro-ph.HE}
}

@article{Patruno2017c,
  author = {{Patruno}, Alessandro and {Jaodand}, Amruta and {Kuiper}, Lucien and {Bult}, Peter and {Hessels}, Jason W.~T. and {Knigge}, Christian and {King}, Andrew R. and {Wijnands}, Rudy and {van der Klis}, Michiel},
  title = {{Radio Pulse Search and X-Ray Monitoring of SAX J1808.4-3658: What Causes Its Orbital Evolution?}},
  journal = {\apj},
  year = {2017},
  month = {jun},
  volume = {841},
  number = {2},
  pages = {98},
  doi = {10.3847/1538-4357/aa6f5b},
  eprint = {1611.06023},
  archivePrefix = {arXiv},
  primaryClass = {astro-ph.HE}
}

@inproceedings{Patruno2021,
  author = {{Patruno}, Alessandro and {Watts}, Anna L.},
  editor = {{Belloni}, Tomaso M. and {M{\'e}ndez}, Mariano and {Zhang}, Chengmin},
  title = {{Accreting Millisecond X-ray Pulsars}},
  booktitle = {Timing Neutron Stars: Pulsations, Oscillations and Explosions},
  series = {Astrophysics and Space Science Library},
  year = {2021},
  month = {jan},
  volume = {461},
  pages = {143--208},
  doi = {10.1007/978-3-662-62110-3_4},
  eprint = {1206.2727},
  archivePrefix = {arXiv},
  primaryClass = {astro-ph.HE}
}

@article{PhysRevD.105.022002,
  author = {Abbott, R. and Abbott, T. D. and Acernese, F. and others},
  title = {Search for continuous gravitational waves from 20 accreting millisecond x-ray pulsars in O3 LIGO data},
  journal = {Phys. Rev. D},
  publisher = {American Physical Society},
  year = {2022{\natexlab{a}}},
  month = {Jan},
  volume = {105},
  pages = {022002},
  doi = {10.1103/PhysRevD.105.022002},
  url = {https://link.aps.org/doi/10.1103/PhysRevD.105.022002}
}

@article{Pike2021_GRS1741,
  author = {Pike, Sean N. and Harrison, Fiona A. and Tomsick, John A. and Bachetti, Matteo and Buisson, Douglas J. K. and Garcia, Javier and Jiang, Jiachen and Ludlam, R. M. and Madsen, Kristin K.},
  title = {Photospheric Radius Expansion and a double-peaked type-I X-ray burst from GRS 1741.9-2853},
  journal = {The Astrophysical Journal},
  year = {2021},
  volume = {918},
  pages = {9},
  doi = {10.3847/1538-4357/ac0ef9},
  eprint = {2106.13312},
  archivePrefix = {arXiv}
}

@article{Podsiadlowski2002,
  author = {{Podsiadlowski}, Ph. and {Rappaport}, S. and {Pfahl}, E.~D.},
  title = {{Evolutionary Sequences for Low- and Intermediate-Mass X-Ray Binaries}},
  journal = {\apj},
  year = {2002},
  month = {feb},
  volume = {565},
  number = {2},
  pages = {1107--1133},
  doi = {10.1086/324686},
  eprint = {astro-ph/0107261},
  archivePrefix = {arXiv},
  primaryClass = {astro-ph}
}

@article{Potekhin1997,
  author = {Potekhin, A. Y. and Chabrier, G. and Yakovlev, D. G.},
  title = {Internal temperatures and cooling of neutron stars with accreted envelopes},
  journal = {Astronomy and Astrophysics},
  year = {1997},
  volume = {323},
  pages = {415--428},
  eprint = {astro-ph/9706148},
  archivePrefix = {arXiv}
}

@article{Potekhin2019,
  author = {Potekhin, A. Y. and Chugunov, A. I. and Chabrier, G.},
  title = {Thermal evolution and quiescent emission of transiently accreting neutron stars},
  journal = {Astronomy and Astrophysics},
  year = {2019},
  volume = {629},
  pages = {A88},
  doi = {10.1051/0004-6361/201936003},
  eprint = {1907.08299},
  archivePrefix = {arXiv},
  primaryClass = {astro-ph.HE}
}

@article{RadhakrishnanSrinivasan1982,
  author = {Radhakrishnan, V. and Srinivasan, G.},
  title = {On the origin of the recently discovered ultra-rapid pulsar},
  journal = {Current Science},
  year = {1982},
  volume = {51},
  number = {23},
  pages = {1096--1099},
  url = {https://repository.ias.ac.in/55563/1/17_pub.pdf}
}

@article{RilesReview,
  author = {{Riles}, Keith},
  title = {{Searches for continuous-wave gravitational radiation}},
  journal = {Living Reviews in Relativity},
  year = {2023},
  month = {dec},
  volume = {26},
  number = {1},
  pages = {3},
  doi = {10.1007/s41114-023-00044-3},
  eprint = {2206.06447},
  archivePrefix = {arXiv},
  primaryClass = {astro-ph.HE}
}

@article{Rutledge2002,
  author = {{Rutledge}, Robert E. and {Bildsten}, Lars and {Brown}, Edward F. and {Pavlov}, George G. and {Zavlin}, Vyacheslav E. and {Ushomirsky}, Greg},
  title = {{Crustal Emission and the Quiescent Spectrum of the Neutron Star in KS 1731-260}},
  journal = {\apj},
  year = {2002},
  month = {nov},
  volume = {580},
  number = {1},
  pages = {413--422},
  doi = {10.1086/342745},
  eprint = {astro-ph/0108125},
  archivePrefix = {arXiv},
  primaryClass = {astro-ph}
}

@article{Sanna2017_MAXIJ0911,
  author = {Sanna, A. and Papitto, A. and Burderi, L. and Bozzo, E. and Riggio, A. and Di Salvo, T. and Ferrigno, C. and Rea, N. and Iaria, R.},
  title = {Discovery of a new accreting millisecond X-ray pulsar in the globular cluster NGC 2808},
  journal = {Astronomy and Astrophysics},
  year = {2017},
  volume = {598},
  pages = {A34},
  doi = {10.1051/0004-6361/201629406},
  eprint = {1611.02995},
  archivePrefix = {arXiv}
}

@article{Sanna2018IGRJ17591,
  author = {Sanna, A. and Ferrigno, C. and Ray, P. S. and Ducci, L. and Jaisawal, G. K. and Enoto, T. and Bozzo, E. and Altamirano, D. and Di Salvo, T. and Strohmayer, T. E. and others},
  title = {NuSTAR and NICER reveal IGR J17591-2342 as a new accreting millisecond X-ray pulsar},
  journal = {Astronomy and Astrophysics},
  year = {2018},
  volume = {617},
  pages = {L8},
  doi = {10.1051/0004-6361/201834160},
  eprint = {1808.10195},
  archivePrefix = {arXiv},
  primaryClass = {astro-ph.HE}
}

@article{Sanna2018_IGRJ16597,
  author = {{Sanna}, A. and {Bahramian}, A. and {Bozzo}, E. and {Heinke}, C. and {Altamirano}, D. and {Wijnands}, R. and {Degenaar}, N. and {Maccarone}, T. and {Riggio}, A. and {Di Salvo}, T. and {Iaria}, R. and {Burgay}, M. and {Possenti}, A. and {Ferrigno}, C. and {Papitto}, A. and {Sivakoff}, G. R. and {D'Amico}, N. and {Burderi}, L.},
  title = {Discovery of 105 Hz coherent pulsations in the ultracompact binary IGR J16597-3704},
  journal = {Astronomy and Astrophysics},
  year = {2018},
  volume = {610},
  pages = {L2},
  doi = {10.1051/0004-6361/201732262},
  eprint = {1711.03092},
  archivePrefix = {arXiv}
}

@article{Sanna2018_IGRJ17379,
  author = {Sanna, A. and Bozzo, E. and Papitto, A. and Riggio, A. and Ferrigno, C. and Di Salvo, T. and Iaria, R. and Mazzola, S. M. and D'Amico, N. and Burderi, L.},
  title = {XMM-Newton detection of the 2.1 ms coherent pulsations from IGR J17379-3747},
  journal = {Astronomy and Astrophysics},
  year = {2018},
  volume = {616},
  pages = {L17},
  doi = {10.1051/0004-6361/201833205},
  eprint = {1807.08574},
  archivePrefix = {arXiv}
}

@article{Sanna2022_MAXIJ1957,
  author = {{Sanna}, A and {Bult}, P and {Ng}, M and {Ray}, P S and {Jaisawal}, G K and {Burderi}, L and {Di Salvo}, T and {Riggio}, A and {Altamirano}, D and {Strohmayer}, T E and {Manca}, A and {Gendreau}, K C and {Chakrabarty}, D and {Iwakiri}, W and {Iaria}, R},
  title = {MAXI J1957+032: a new accreting millisecond X-ray pulsar in an ultra-compact binary},
  journal = {Monthly Notices of the Royal Astronomical Society: Letters},
  year = {2022},
  volume = {516},
  pages = {L76--L80},
  doi = {10.1093/mnrasl/slac093},
  eprint = {2208.05807},
  archivePrefix = {arXiv}
}

@article{Sanna2022b,
  author = {{Sanna}, A. and {Burderi}, L. and {Di Salvo}, T. and {Riggio}, A. and {Altamirano}, D. and others},
  title = {{On the peculiar long-term orbital evolution of the eclipsing accreting millisecond X-ray pulsar SWIFT J1749.4--2807}},
  journal = {Monthly Notices of the Royal Astronomical Society},
  year = {2022},
  volume = {514},
  number = {3},
  pages = {4385--4397},
  doi = {10.1093/mnras/stac1611}
}

@article{ShaoLi2012,
  author = {Shao, Yong and Li, Xiang-Dong},
  title = {Formation of Millisecond Pulsars from Intermediate- and Low-Mass X-ray Binaries},
  journal = {The Astrophysical Journal},
  year = {2012},
  volume = {756},
  pages = {85},
  doi = {10.1088/0004-637X/756/1/85},
  eprint = {1207.2833},
  archivePrefix = {arXiv},
  primaryClass = {astro-ph.SR}
}

@article{SidoliSguera2024_16470,
  author = {Sidoli, L. and Sguera, V.},
  title = {{Archival X-ray observations of SRGA J144459.2--604207 and a candidate 2MASS counterpart}},
  journal = {The Astronomer's Telegram},
  year = {2024},
  volume = {16470},
  pages = {1},
  url = {https://www.astronomerstelegram.org/?read=16470}
}

@article{Singh2025,
  author = {{Singh}, A. and {Sanna}, A. and {Bhattacharyya}, S. and {Chakraborty}, S. and {Jangle}, S. and {Katoch}, T. and {Antia}, H.~M. and {Bijewar}, N.},
  title = {{AstroSat timing and spectral analysis of the accretion-powered millisecond X-ray pulsar IGR J17591--2342}},
  journal = {Monthly Notices of the Royal Astronomical Society},
  year = {2025},
  volume = {536},
  number = {2},
  pages = {1323--1331},
  doi = {10.1093/mnras/stae2640}
}

@article{Strohmayer2017_IGRJ17062,
  author = {Strohmayer, Tod E. and Keek, Laurens},
  title = {IGR J17062-6143 is an Accreting Millisecond X-ray Pulsar},
  journal = {The Astrophysical Journal Letters},
  year = {2017},
  volume = {836},
  pages = {L23},
  doi = {10.3847/2041-8213/aa5e51},
  eprint = {1702.05449},
  archivePrefix = {arXiv}
}

@article{Tauris2012,
  author = {{Tauris}, Thomas M.},
  title = {{Spin-Down of Radio Millisecond Pulsars at Genesis}},
  journal = {Science},
  year = {2012},
  month = {feb},
  volume = {335},
  number = {6068},
  pages = {561--563},
  doi = {10.1126/science.1216355},
  eprint = {1202.0551},
  archivePrefix = {arXiv},
  primaryClass = {astro-ph.SR}
}

@article{Wachter2002,
  author = {Wachter, S. and Hoard, D. W. and Bailyn, C. D. and Corbel, S. and Kaaret, P.},
  title = {A Closer Look at the Soft X-Ray Transient X1608--52: Long-Term Optical and X-Ray Observations},
  journal = {\apj},
  year = {2002},
  volume = {568},
  pages = {901--911},
  doi = {10.1086/339034},
  eprint = {astro-ph/0112465},
  archivePrefix = {arXiv}
}

@article{Wang1995,
  author = {Wang, Y.-M.},
  title = {On the torque exerted by a magnetically threaded accretion disk},
  journal = {The Astrophysical Journal Letters},
  year = {1995},
  volume = {449},
  pages = {L153--L156},
  doi = {10.1086/309649}
}

@article{WangLagakosGray2010,
  author = {Wang, Rui and Lagakos, Stephen W. and Gray, Robert J.},
  title = {Testing and interval estimation for two-sample survival comparisons with small sample sizes and unequal censoring},
  journal = {Biostatistics},
  year = {2010},
  volume = {11},
  number = {4},
  pages = {676--692},
  doi = {10.1093/biostatistics/kxq021}
}

@article{Watts2008GW,
  author = {Watts, A. L. and Krishnan, B. and Bildsten, L. and Schutz, B. F.},
  title = {Detecting gravitational wave emission from the known accreting neutron stars},
  journal = {Monthly Notices of the Royal Astronomical Society},
  year = {2008},
  volume = {389},
  pages = {839--868},
  doi = {10.1111/j.1365-2966.2008.13594.x},
  eprint = {0803.4097},
  archivePrefix = {arXiv}
}

@article{WhiteZhang1997,
  author = {White, N. E. and Zhang, W.},
  title = {Millisecond X-ray pulsars in low-mass X-ray binaries},
  journal = {The Astrophysical Journal Letters},
  year = {1997},
  volume = {490},
  pages = {L87--L90},
  doi = {10.1086/311018}
}

@article{Wijnands2013,
  author = {{Wijnands}, R. and {Degenaar}, N. and {Page}, D.},
  title = {{Testing the deep-crustal heating model using quiescent neutron-star very-faint X-ray transients and the possibility of partially accreted crusts in accreting neutron stars}},
  journal = {\mnras},
  year = {2013},
  month = {jul},
  volume = {432},
  number = {3},
  pages = {2366--2377},
  doi = {10.1093/mnras/stt599},
  eprint = {1208.4273},
  archivePrefix = {arXiv},
  primaryClass = {astro-ph.HE}
}

@article{Yakovlev2004,
  author = {Yakovlev, D. G. and Levenfish, K. P. and Potekhin, A. Y. and Gnedin, O. Y. and Chabrier, G.},
  title = {Thermal states of coldest and hottest neutron stars in soft X-ray transients},
  journal = {Astronomy and Astrophysics},
  year = {2004},
  volume = {417},
  pages = {169--179},
  doi = {10.1051/0004-6361:20034191},
  eprint = {astro-ph/0310259},
  archivePrefix = {arXiv}
}

@article{Yakovlev2004_rev,
  author = {Yakovlev, D. G. and Pethick, C. J.},
  title = {Neutron Star Cooling},
  journal = {Annu. Rev. Astron. Astrophys.},
  year = {2004},
  volume = {42},
  pages = {169--210},
  doi = {10.1146/annurev.astro.42.053102.134013}
}

@article{YangLi2024,
  author = {{Yang}, Hao-Ran and {Li}, Xiang-Dong},
  title = {{The Influence of the Magnetic Braking Laws on the Evolution of Persistent and Transient Low-mass X-Ray Binaries}},
  journal = {\apj},
  year = {2024},
  month = {oct},
  volume = {974},
  number = {2},
  pages = {298},
  doi = {10.3847/1538-4357/ad7824},
  eprint = {2409.05067},
  archivePrefix = {arXiv},
  primaryClass = {astro-ph.HE}
}

@article{cavecchi2026,
  author = {{Cavecchi}, Y. and {Galloway}, D.~K. and {Heger}, A. and {Santill{\'a}n-Ortega}, P. and {Nava-Callejas}, M. and {Vincentelli}, F.~M. and {Rivera Sandoval}, L.~E. and {Goodwin}, A. and {Johnston}, Z. and {Puente Mancilla}, S. and {Page}, D.},
  title = {{A Solution to the Tension of Burning on Neutron Stars and Nuclear Physics}},
  journal = {\apj},
  year = {2026},
  month = {feb},
  volume = {998},
  number = {1},
  pages = {25},
  doi = {10.3847/1538-4357/ae2ff3},
  eprint = {2506.11966},
  archivePrefix = {arXiv},
  primaryClass = {astro-ph.HE}
}

@article{gusakov2014,
  author = {{Gusakov}, Mikhail E. and {Chugunov}, Andrey I. and {Kantor}, Elena M.},
  title = {{Instability Windows and Evolution of Rapidly Rotating Neutron Stars}},
  journal = {\prl},
  year = {2014},
  month = {apr},
  volume = {112},
  number = {15},
  pages = {151101},
  doi = {10.1103/PhysRevLett.112.151101},
  eprint = {1310.8103},
  archivePrefix = {arXiv},
  primaryClass = {astro-ph.HE}
}

@article{nava-callejas2026,
  author = {{Nava-Callejas}, Martin and {Cavecchi}, Yuri and {Page}, Dany},
  title = {{On a re-examination of neutron star cooling in transient sources -- No shallow heating required?}},
  journal = {arXiv e-prints},
  year = {2026},
  month = {jun},
  pages = {arXiv:2606.02932},
  eprint = {2606.02932},
  archivePrefix = {arXiv},
  primaryClass = {astro-ph.HE},
  url = {https://arxiv.org/abs/2606.02932}
}

@article{niang2026,
  author = {{Niang}, N. and {Ertan}, {\"U}. and {Gen{\c{c}}ali}, A.~A. and {Ertu{\u{g}}rul}, F. and {Ulubay}, A. and {Devlen}, E. and {Alpar}, M.~A.},
  title = {{Binary and neutron star evolution in low-mass X-ray binaries on the evolutionary tracks of accreting millisecond X-ray pulsars}},
  journal = {\mnras},
  year = {2026},
  month = {may},
  volume = {548},
  number = {3},
  pages = {stag745},
  doi = {10.1093/mnras/stag745},
  eprint = {2604.16036},
  archivePrefix = {arXiv},
  primaryClass = {astro-ph.HE}
}

@article{riverasandoval2020,
  author = {{Rivera Sandoval}, L.~E. and {Maccarone}, T.~J. and {Pichardo Marcano}, M.},
  title = {{A Year-long Superoutburst from an Ultracompact White Dwarf Binary Reveals the Importance of Donor Star Irradiation}},
  journal = {\apjl},
  year = {2020},
  month = {sep},
  volume = {900},
  number = {2},
  pages = {L37},
  doi = {10.3847/2041-8213/abb130},
  eprint = {2008.09242},
  archivePrefix = {arXiv},
  primaryClass = {astro-ph.HE}
}

@article{sugiura1978,
  author = {Sugiura, Nariaki},
  title = {Further analysis of the data by Akaike's information criterion and the finite corrections},
  journal = {Communications in Statistics - Theory and Methods},
  year = {1978},
  volume = {7},
  number = {1},
  pages = {13--26},
  doi = {10.1080/03610927808827599}
}

@article{thermalBahramian2015,
  author = {Bahramian, Arash and Heinke, Craig O. and Degenaar, Nathalie and Chomiuk, Laura and Wijnands, Rudy and Strader, Jay and Ho, Wynn C. G. and Pooley, David},
  title = {Limits on thermal variations in a dozen quiescent neutron stars over a decade},
  journal = {Monthly Notices of the Royal Astronomical Society},
  year = {2015},
  volume = {452},
  number = {4},
  pages = {3475--3488},
  doi = {10.1093/mnras/stv1585},
  eprint = {1507.03994},
  archivePrefix = {arXiv},
  url = {https://academic.oup.com/mnras/article/452/4/3475/1063875}
}

@article{thermalDegenaar2012,
  author = {Degenaar, N. and Patruno, A. and Wijnands, R.},
  title = {The Quiescent {X-Ray} Properties of the Accreting Millisecond {X-Ray} Pulsar and Eclipsing Binary {Swift J1749.4-2807}},
  journal = {The Astrophysical Journal},
  year = {2012},
  volume = {756},
  number = {2},
  pages = {148},
  doi = {10.1088/0004-637X/756/2/148},
  eprint = {1204.6059},
  archivePrefix = {arXiv},
  url = {https://arxiv.org/abs/1204.6059}
}

@article{thermalDegenaar2021,
  author = {Degenaar, N. and Page, D. and van den Eijnden, J. and Beznogov, M. V. and Wijnands, R. and Reynolds, M.},
  title = {Constraining the properties of dense neutron star cores: the case of the low-mass {X-ray} binary {HETE J1900.1-2455}},
  journal = {Monthly Notices of the Royal Astronomical Society},
  year = {2021},
  volume = {508},
  pages = {882--894},
  doi = {10.1093/mnras/stab2202},
  url = {https://academic.oup.com/mnras/article/508/1/882/6339290}
}

@article{thermalHaskell2012,
  author = {Haskell, B. and Degenaar, N. and Ho, W. C. G.},
  title = {Constraining the physics of the r-mode instability in neutron stars with X-ray and ultraviolet observations},
  journal = {Monthly Notices of the Royal Astronomical Society},
  year = {2012},
  volume = {424},
  pages = {93--103},
  doi = {10.1111/j.1365-2966.2012.21171.x},
  url = {https://arxiv.org/abs/1201.2101}
}

@article{thermalHeinke2009,
  author = {Heinke, C. O. and Jonker, P. G. and Wijnands, R. and Deloye, C. J. and Taam, R. E.},
  title = {Further Constraints on Thermal Quiescent {X-Ray} Emission from {SAX J1808.4-3658}},
  journal = {The Astrophysical Journal},
  year = {2009},
  volume = {691},
  number = {2},
  pages = {1035--1041},
  doi = {10.1088/0004-637X/691/2/1035},
  eprint = {0810.0497},
  archivePrefix = {arXiv},
  url = {https://arxiv.org/abs/0810.0497}
}

@article{thermalHoman2016,
  author = {Homan, Jeroen and Sivakoff, Gregory and Pooley, David and Heinke, Craig and Strader, Jay and Tudor, Vlad and Miller-Jones, James},
  title = {{Chandra} identification of the X-ray transient {MAXI J0911-635/Swift J0911.9-6452} in {NGC 2808}},
  journal = {The Astronomer's Telegram},
  year = {2016},
  volume = {8971},
  pages = {1},
  url = {https://www.astronomerstelegram.org/?read=8971}
}

@article{thermalJonker2004,
  author = {Jonker, P. G. and Wijnands, R. and van der Klis, M.},
  title = {The faint neutron star soft X-ray transient {SAX J1810.8-2609} in quiescence},
  journal = {Monthly Notices of the Royal Astronomical Society},
  year = {2004},
  volume = {349},
  pages = {94--98},
  doi = {10.1111/j.1365-2966.2004.07455.x},
  url = {https://arxiv.org/abs/astro-ph/0311560}
}

@article{thermalLinares2014,
  author = {Linares, Manuel and Bahramian, Arash and Heinke, Craig and Wijnands, Rudy and Patruno, Alessandro and Altamirano, Diego and Homan, Jeroen and Bogdanov, Slavko and Pooley, David},
  title = {The neutron star transient and millisecond pulsar in {M28}: from sub-luminous accretion to rotation-powered quiescence},
  journal = {Monthly Notices of the Royal Astronomical Society},
  year = {2014},
  volume = {438},
  pages = {251--261},
  doi = {10.1093/mnras/stt2167},
  url = {https://arxiv.org/abs/1310.7937}
}

@article{thermalMarino2022,
  author = {Marino, A. and Russell, T. D. and Del Santo, M. and Coti Zelati, F. and Rea, N. and Manca, A. and Sanna, A.},
  title = {Radio and X-ray observations of the {AMXP MAXI J1816-195} in quiescence},
  journal = {The Astronomer's Telegram},
  year = {2022},
  volume = {15769},
  pages = {1},
  url = {https://www.astronomerstelegram.org/?read=15769}
}

@article{thermalMataSanchez2017,
  author = {{Mata S{\'a}nchez}, D. and Charles, P. A. and Armas Padilla, M. and Buckley, D. A. H. and Israel, G. L. and Linares, M. and Mu{\~n}oz-Darias, T.},
  title = {{Swift} and {SALT} observations of the multiple outbursts of {MAXI J1957+032}},
  journal = {Monthly Notices of the Royal Astronomical Society},
  year = {2017},
  volume = {468},
  pages = {564--569},
  doi = {10.1093/mnras/stx479},
  url = {https://arxiv.org/abs/1701.03381}
}

@article{thermalMerritt2016,
  author = {Merritt, Rachael L. and Cackett, Edward M. and Brown, Edward F. and Page, Dany and Cumming, Andrew and Degenaar, Nathalie and Deibel, Alex and Homan, Jeroen and Miller, Jon M. and Wijnands, Rudy},
  title = {The Thermal State of {KS 1731-260} after 14.5 Years in Quiescence},
  journal = {The Astrophysical Journal},
  year = {2016},
  volume = {833},
  number = {2},
  pages = {186},
  doi = {10.3847/1538-4357/833/2/186},
  eprint = {1608.03880},
  archivePrefix = {arXiv},
  url = {https://arxiv.org/abs/1608.03880}
}

@article{thermalMorrison1983,
  author = {Morrison, Robert and McCammon, Dan},
  title = {Interstellar photoelectric absorption cross sections, 0.03--10 keV},
  journal = {The Astrophysical Journal},
  year = {1983},
  volume = {270},
  pages = {119--122},
  doi = {10.1086/161102},
  url = {https://articles.adsabs.harvard.edu/pdf/1983ApJ...270..119M}
}

@article{thermalMuno2003,
  author = {Muno, M. P. and Baganoff, F. K. and Arabadjis, J. S.},
  title = {The X-Ray Binary {GRS 1741.9-2853} in Outburst and Quiescence},
  journal = {The Astrophysical Journal},
  year = {2003},
  volume = {598},
  pages = {474--480},
  doi = {10.1086/378851},
  url = {https://arxiv.org/abs/astro-ph/0307179}
}

@article{thermalParikh2017,
  author = {Parikh, A. S. and Wijnands, R.},
  title = {Variable Quiescent State for the Neutron-Star {X-Ray} Transient {SAX J1750.8-2900}: Not Such a Hot Neutron Star after All?},
  journal = {Monthly Notices of the Royal Astronomical Society},
  year = {2017},
  volume = {472},
  number = {3},
  pages = {2742--2750},
  doi = {10.1093/mnras/stx2154},
  url = {https://academic.oup.com/mnras/article/472/3/2742/4085644}
}

@article{thermalParikh2019,
  author = {Parikh, A. S. and Wijnands, R. and Ootes, L. S. and Page, D. and Degenaar, N. and Bahramian, A. and Brown, E. F. and Cackett, E. M. and Cumming, A. and Heinke, C. and Homan, J. and Rouco Escorial, A. and Wijngaarden, M. J. P.},
  title = {Consistent accretion-induced heating of the neutron-star crust in {MXB 1659-29} during two different outbursts},
  journal = {Astronomy \& Astrophysics},
  year = {2019},
  volume = {624},
  pages = {A84},
  doi = {10.1051/0004-6361/201834412},
  eprint = {1810.05626},
  archivePrefix = {arXiv}
}

@article{thermalParikh2020,
  author = {Parikh, A. S. and Wijnands, R. and Homan, J. and Degenaar, N. and Wolvers, B. and Ootes, L. S. and Page, D.},
  title = {Unexpected late-time temperature increase observed in the two neutron star crust-cooling sources {XTE J1701-462} and {EXO 0748-676}},
  journal = {Astronomy \& Astrophysics},
  year = {2020},
  volume = {638},
  pages = {L2},
  doi = {10.1051/0004-6361/202038198},
  eprint = {2005.12014},
  archivePrefix = {arXiv}
}

@article{thermalRavi2017,
  author = {Ravi, Vikram},
  title = {{MAXI J1957+032}: An Accreting Neutron Star Possibly in a Triple System},
  journal = {The Astrophysical Journal},
  year = {2017},
  volume = {851},
  pages = {114},
  doi = {10.3847/1538-4357/aa9277},
  url = {https://arxiv.org/abs/1710.01333}
}

@article{thermalRutledge1999,
  author = {Rutledge, Robert E. and Bildsten, Lars and Brown, Edward F. and Pavlov, George G. and Zavlin, Vyacheslav E.},
  title = {The Thermal X-Ray Spectra of {Centaurus X-4}, {Aquila X-1}, and {4U 1608-522} in Quiescence},
  journal = {The Astrophysical Journal},
  year = {1999},
  volume = {514},
  pages = {945--951},
  doi = {10.1086/306990},
  url = {https://arxiv.org/abs/astro-ph/9810288}
}

@article{thermalServillat2008,
  author = {Servillat, M. and Dieball, A. and Webb, N. A. and Knigge, C. and Cornelisse, R. and Barret, D. and Long, K. S. and Shara, M. M. and Zurek, D. R.},
  title = {Combined {Chandra}, {XMM-Newton}, and {Hubble Space Telescope} observations of the Galactic globular cluster {NGC 2808}},
  journal = {Astronomy and Astrophysics},
  year = {2008},
  volume = {490},
  pages = {641--654},
  doi = {10.1051/0004-6361:200810188},
  url = {https://arxiv.org/abs/0808.2661}
}

@article{thermalTetarenko2018,
  author = {Tetarenko, A. J. and Bahramian, A. and Wijnands, R. and others},
  title = {A Radio Frequency Study of the Accreting Millisecond X-Ray Pulsar, {IGR J16597-3704}, in the Globular Cluster {NGC 6256}},
  journal = {The Astrophysical Journal},
  year = {2018},
  volume = {854},
  pages = {125},
  doi = {10.3847/1538-4357/aaa95a},
  url = {https://arxiv.org/abs/1801.05778}
}

@article{thermalTomsick2007,
  author = {Tomsick, John A. and Gelino, Dawn M. and Kaaret, Philip},
  title = {Uncovering the Nature of the X-Ray Transient {4U 1730-22}: Discovery of X-Ray Emission from a Neutron Star in Quiescence with {Chandra}},
  journal = {The Astrophysical Journal},
  year = {2007},
  volume = {663},
  pages = {461--467},
  doi = {10.1086/518239},
  url = {https://arxiv.org/abs/astro-ph/0703501}
}

@ARTICLE{2009ApJ...698L..60P,
       author = {{Patruno}, Alessandro and {Wijnands}, Rudy and {van der Klis}, Michiel},
        title = "{An Alternative Interpretation of the Timing Noise in Accreting Millisecond Pulsars}",
      journal = {\apjl},
         year = 2009,
        month = jun,
       volume = {698},
       number = {1},
        pages = {L60-L63},
          doi = {10.1088/0004-637X/698/1/L60},
archivePrefix = {arXiv},
       eprint = {0902.4118},
 primaryClass = {astro-ph.HE},
       adsurl = {https://ui.adsabs.harvard.edu/abs/2009ApJ...698L..60P}
}

@article{mann1947,
  author = {{Mann}, H.~B. and {Whitney}, D.~R.},
  title = {{On a Test of Whether One of Two Random Variables Is Stochastically Larger than the Other}},
  journal = {Annals of Mathematical Statistics},
  year = {1947},
  volume = {18},
  number = {1},
  pages = {50--60},
  doi = {10.1214/aoms/1177730491}
}

@ARTICLE{oleary2024b,
       author = {{O'Leary}, Joseph and {Melatos}, Andrew and {Kimpson}, Tom and {O'Neill}, Nicholas J. and {Meyers}, Patrick M. and {Christodoulou}, Dimitris M. and {Bhattacharya}, Sayantan and {Laycock}, Silas G.~T.},
        title = "{Measuring the Magnetic Dipole Moment and Magnetospheric Fluctuations of Accretion-powered Pulsars in the Small Magellanic Cloud with an Unscented Kalman Filter}",
      journal = {\apj},
         year = 2024,
        month = aug,
       volume = {971},
       number = {2},
          eid = {126},
        pages = {126},
          doi = {10.3847/1538-4357/ad53c2},
archivePrefix = {arXiv},
       eprint = {2406.01827},
 primaryClass = {astro-ph.HE},
       adsurl = {https://ui.adsabs.harvard.edu/abs/2024ApJ...971..126O}
}

@ARTICLE{oleary2024,
       author = {{O'Leary}, Joseph and {Melatos}, Andrew and {O'Neill}, Nicholas J. and {Meyers}, Patrick M. and {Christodoulou}, Dimitris M. and {Bhattacharya}, Sayantan and {Laycock}, Silas G.~T.},
        title = "{Measuring the Magnetic Dipole Moment and Magnetospheric Fluctuations of SXP 18.3 with a Kalman Filter}",
      journal = {\apj},
         year = 2024,
        month = apr,
       volume = {965},
       number = {2},
          eid = {102},
        pages = {102},
          doi = {10.3847/1538-4357/ad2adc},
archivePrefix = {arXiv},
       eprint = {2402.11991},
 primaryClass = {astro-ph.HE},
       adsurl = {https://ui.adsabs.harvard.edu/abs/2024ApJ...965..102O}
}

@article{thermalWaterhouse2016,
  author = {Waterhouse, A. C. and Degenaar, N. and Wijnands, R. and Brown, E. F. and Miller, J. M. and Altamirano, D. and Linares, M.},
  title = {Constraining the properties of neutron star crusts with the transient low-mass {X-ray} binary {Aql X-1}},
  journal = {Monthly Notices of the Royal Astronomical Society},
  year = {2016},
  volume = {456},
  pages = {4001--4014},
  doi = {10.1093/mnras/stv2959},
  url = {https://academic.oup.com/mnras/article/456/4/4001/2892766}
}

@article{thermalWijnands2005,
  author = {Wijnands, Rudy and Homan, Jeroen and Heinke, Craig O. and Miller, Jon M. and Lewin, Walter H. G.},
  title = {{Chandra} Observations of the Accretion-driven Millisecond {X-Ray} Pulsars {XTE J0929-314} and {XTE J1751-305} in Quiescence},
  journal = {The Astrophysical Journal},
  year = {2005},
  volume = {619},
  number = {1},
  pages = {492--502},
  doi = {10.1086/426379},
  eprint = {astro-ph/0406057},
  archivePrefix = {arXiv},
  url = {https://arxiv.org/abs/astro-ph/0406057}
}

@article{LIGOScientific:2014pky,
    author = "Aasi, J. and others",
    collaboration = "LIGO Scientific",
    title = "{Advanced LIGO}",
    eprint = "1411.4547",
    archivePrefix = "arXiv",
    primaryClass = "gr-qc",
    doi = "10.1088/0264-9381/32/7/074001",
    journal = "Class. Quant. Grav.",
    volume = "32",
    pages = "074001",
    year = "2015"
}

@article{VIRGO:2014yos,
    author = "Acernese, F. and others",
    collaboration = "VIRGO",
    title = "{Advanced Virgo: a second-generation interferometric gravitational wave detector}",
    eprint = "1408.3978",
    archivePrefix = "arXiv",
    primaryClass = "gr-qc",
    doi = "10.1088/0264-9381/32/2/024001",
    journal = "Class. Quant. Grav.",
    volume = "32",
    number = "2",
    pages = "024001",
    year = "2015"
}

@article{KAGRA:2020tym,
    author = "Akutsu, T. and others",
    collaboration = "KAGRA",
    title = "{Overview of KAGRA: Detector design and construction history}",
    eprint = "2005.05574",
    archivePrefix = "arXiv",
    primaryClass = "physics.ins-det",
    doi = "10.1093/ptep/ptaa125",
    journal = "PTEP",
    volume = "2021",
    number = "5",
    pages = "05A101",
    year = "2021"
}

@article{LIGOScientific:2026ppd,
    author = "Abac, A. G. and others",
    collaboration = "LIGO Scientific, Virgo, KAGRA",
    title = "{Sub-Torque-Balance Upper Limits on Continuous Gravitational Waves from Scorpius X-1}",
    journal = "arXiv e-prints",
    eprint = "2607.07765",
    archivePrefix = "arXiv",
    pages = "arXiv:2607.07765",
    primaryClass = "astro-ph.HE",
    year = "2026"
}

@article{Whelan:2023bha,
    author = "Whelan, John T. and others",
    title = "{Search for Gravitational Waves from Scorpius X-1 in LIGO O3 Data with Corrected Orbital Ephemeris}",
    eprint = "2302.10338",
    archivePrefix = "arXiv",
    primaryClass = "astro-ph.HE",
    reportNumber = "LIGO-P2300042-v6",
    doi = "10.3847/1538-4357/acc8d7",
    journal = "Astrophys. J.",
    volume = "949",
    number = "2",
    pages = "117",
    year = "2023"
}

@article{antonelliGravitationalPulsarsCorrelations2025,
  title = {Gravitational Pulsars: Correlations between the Electromagnetic and the Continuous Gravitational Wave Signal},
  shorttitle = {Gravitational Pulsars},
  author = {Antonelli, Marco and Basu, Avishek and Haskell, Brynmor},
  year = 2025,
  journal = {Publications of the Astronomical Society of Australia},
  volume = {42},
  eprint = {2507.18439},
  primaryclass = {astro-ph},
  pages = {e118},
  issn = {1323-3580, 1448-6083},
  doi = {10.1017/pasa.2025.10086},
  urldate = {2025-11-19},
  archiveprefix = {arXiv},
}

@ARTICLE{Melatos23,
       author = {{Melatos}, A. and {O'Neill}, N.~J. and {Meyers}, P.~M. and {O'Leary}, J.},
        title = "{Tracking Hidden Magnetospheric Fluctuations in Accretion-powered Pulsars With a Kalman Filter}",
      journal = {\apj},
         year = 2023,
        month = feb,
       volume = {944},
       number = {1},
          eid = {64},
        pages = {64},
          doi = {10.3847/1538-4357/acab5a},
archivePrefix = {arXiv},
       eprint = {2212.08185},
 primaryClass = {astro-ph.HE},
       adsurl = {https://ui.adsabs.harvard.edu/abs/2023ApJ...944...64M}
}

@ARTICLE{Melatos26,
       author = {{Melatos}, Andrew},
        title = "{Spin wandering of an accreting neutron star near torque balance: implications for gravitational wave searches}",
      journal = {arXiv e-prints},
         year = 2026,
        month = sep,
          eid = {arXiv:2609.32164},
        pages = {arXiv:2609.32164},
          doi = {10.48550/arXiv.2609.32164},
archivePrefix = {arXiv},
       eprint = {2609.32164},
 primaryClass = {astro-ph.HE},
       adsurl = {https://ui.adsabs.harvard.edu/abs/2026arXiv260932164M}
}

@ARTICLE{2023APh...15302880W,
       author = {{Wette}, Karl},
        title = "{Searches for continuous gravitational waves from neutron stars: A twenty-year retrospective}",
      journal = {Astroparticle Physics},
         year = 2023,
        month = nov,
       volume = {153},
          eid = {102880},
        pages = {102880},
          doi = {10.1016/j.astropartphys.2023.102880},
archivePrefix = {arXiv},
       eprint = {2305.07106},
 primaryClass = {gr-qc},
       adsurl = {https://ui.adsabs.harvard.edu/abs/2023APh...15302880W}
}

@ARTICLE{2020PhRvD.102b3006M,
       author = {{Middleton}, Hannah and {Clearwater}, Patrick and {Melatos}, Andrew and {Dunn}, Liam},
        title = "{Search for gravitational waves from five low mass x-ray binaries in the second Advanced LIGO observing run with an improved hidden Markov model}",
      journal = {\prd},
         year = 2020,
        month = jul,
       volume = {102},
       number = {2},
          eid = {023006},
        pages = {023006},
          doi = {10.1103/PhysRevD.102.023006},
archivePrefix = {arXiv},
       eprint = {2006.06907},
 primaryClass = {astro-ph.HE},
       adsurl = {https://ui.adsabs.harvard.edu/abs/2020PhRvD.102b3006M}
}

@ARTICLE{2017PhRvD..95d2005M,
       author = {{Meadors}, G.~D. and {Goetz}, E. and {Riles}, K. and {Creighton}, T. and {Robinet}, F.},
        title = "{Searches for continuous gravitational waves from Scorpius X-1 and XTE J1751-305 in LIGO's sixth science run}",
      journal = {\prd},
         year = 2017,
        month = feb,
       volume = {95},
       number = {4},
          eid = {042005},
        pages = {042005},
          doi = {10.1103/PhysRevD.95.042005},
archivePrefix = {arXiv},
       eprint = {1610.09391},
 primaryClass = {gr-qc},
       adsurl = {https://ui.adsabs.harvard.edu/abs/2017PhRvD..95d2005M}
}

@ARTICLE{2015PhRvD..92b3006M,
       author = {{Messenger}, C. and {Bulten}, H.~J. and {Crowder}, S.~G. and {Dergachev}, V. and {Galloway}, D.~K. and {Goetz}, E. and {Jonker}, R.~J.~G. and {Lasky}, P.~D. and {Meadors}, G.~D. and {Melatos}, A. and {Premachandra}, S. and {Riles}, K. and {Sammut}, L. and {Thrane}, E.~H. and {Whelan}, J.~T. and {Zhang}, Y.},
        title = "{Gravitational waves from Scorpius X-1: A comparison of search methods and prospects for detection with advanced detectors}",
      journal = {\prd},
         year = 2015,
        month = jul,
       volume = {92},
       number = {2},
          eid = {023006},
        pages = {023006},
          doi = {10.1103/PhysRevD.92.023006},
archivePrefix = {arXiv},
       eprint = {1504.05889},
 primaryClass = {gr-qc},
       adsurl = {https://ui.adsabs.harvard.edu/abs/2015PhRvD..92b3006M}
}
\clearpage
\onecolumn
\begin{appendix}

\section{Spin and orbital properties of the source sample}
\label{app:spin_catalogue}
\begin{table}[ht!]
\centering
\small
\caption{Type, spin frequencies, and orbital periods of accreting neutron
stars in our sample, sorted by orbital period. Spin frequencies and orbital
periods are taken from the discovery or phase-coherent timing papers listed
in the table notes.}
\label{tab:spin_table}
\begin{tabular}{lccc}
\toprule
Name & Type & Frequency (Hz) & Orbit (h) \\
\midrule
IGR\,J17062-6143      & AMXP & 164 & 0.6 \\
XTE\,J1751-305        & AMXP & 435 & 0.7 \\
XTE\,J0929-314        & AMXP & 185 & 0.7 \\
XTE\,J1807-294        & AMXP & 190 & 0.7 \\
MAXI\,J0911-655       & AMXP & 340 & 0.7 \\
IGR\,J16597-3704      & AMXP & 105 & 0.8 \\
Swift\,J1756.9-2508    & AMXP & 182 & 0.9 \\
NGC\,6440X-2          & AMXP & 206 & 0.95 \\
MAXI\,J1957+032       & AMXP & 314 & 1.0 \\
IGR\,J17494-3030      & AMXP & 376 & 1.2 \\
HETE\,J1900.1-2455    & AMXP & 377 & 1.4 \\
IGR\,J17379-3747      & AMXP & 468 & 1.9 \\
SAX\,J1808.4-3658     & AMXP & 401 & 2.0 \\
IGR\,J00291+5934      & AMXP & 599 & 2.45 \\
IGR\,J17511-3057      & AMXP & 245 & 3.5 \\
IGR\,J17498-2921      & AMXP & 401 & 3.8 \\
EXO\,0748-676         & NXP  & 552 & 3.8 \\
4U\,1636-536          & NXP  & 581 & 3.8 \\
XTE\,J1814-338        & AMXP & 314 & 4.3 \\
MAXI\,J1816-195       & AMXP & 528 & 4.8 \\
PSR\,J1023+0038       & AMXP & 592 & 4.8 \\
SRGA\,J144459.2-604207 & AMXP & 448 & 5.2 \\
XSS\,J12270-4859      & AMXP & 593 & 6.9 \\
MXB\,1659-298         & NXP  & 567 & 7.1 \\
SAX\,J1748.9-2021     & AMXP & 442 & 8.8 \\
Swift\,J1749.4-2807   & AMXP & 518 & 8.8 \\
IGR\,J17591-2342      & AMXP & 527 & 8.8 \\
IGR\,J18245-2452      & AMXP & 254 & 11.0 \\
4U\,1608-52$^{*}$     & NXP  & 619 & 12.9 \\
Aql\,X–1              & AMXP & 550 & 18.9 \\
4U\,1728-34           & NXP  & 363 & --- \\
KS\,1731-260          & NXP  & 524 & --- \\
IGR\,J17191-2821      & NXP  & 294 & --- \\
4U\,1702–429          & NXP  & 329 & --- \\
SAX\,J1750.8-2900     & NXP  & 601 & --- \\
GRS\,1741.9-2853      & NXP  & 589 & --- \\
4U\,1730-22           & NXP  & 585 & --- \\
SAX\,J1810.8-2609     & NXP  & 532 & --- \\
XTE\,J1739-285        & AMXP & 386 & --- \\
\bottomrule
\end{tabular}
\tablebib{
Several spins and orbits are taken from \citet{2017ApJ...850..106P}. We cite the corresponding reference when missing in that work, or when an updated value exists. 
IGR~J17062--6143: spin and ultracompact orbit from \citet{Strohmayer2017_IGRJ17062,Bult2021_IGRJ17062};
MAXI~J0911--655: spin and orbit from \citet{Sanna2017_MAXIJ0911};
IGR~J16597--3704: spin and orbit from \citet{Sanna2018_IGRJ16597};
MAXI~J1957+032: spin and orbit from \citet{Sanna2022_MAXIJ1957};
IGR~J17494--3030: spin and orbit from \citet{Ng2021_IGRJ17494};
IGR~J17379--3747: spin and orbit from \citet{Sanna2018_IGRJ17379,Bult2019_IGRJ17379};
MAXI~J1816--195: spin and orbit from \citet{Bult2022_MAXIJ1816};
PSR~J1023+0038: spin and orbit from \citet{Archibald2009_J1023};
IGR~J17591--2342: spin and orbit from \citet{Sanna2018IGRJ17591,Singh2025};
GRS~1741.9--2853: burst properties and spin constraints discussed by \citet{Pike2021_GRS1741};
SRGA~J144459.2--604207: discovery and AMXP identification from \citet{Molkov2024_SRGAJ144459,Ng2024_SRGA_thermal};
XTE~J1739--285: spin constraint and recent outburst properties from \citet{Beri2023_XTEJ1739}.\\
$^{*}$ The orbital period of 4U 1608-52 is determined indirectly from a superhump with a high degree of confidence \citep{Wachter2002}. However, it has never been measured directly, so it is listed here for completeness even if it is not included (as a conservative choice) in the final analysis.    
}
\end{table}
\clearpage

\section{Spin Distribution Models}
\label{app:spin_dist}
\begin{table}[ht!]
\centering
\small
\caption{Comparison of the five fitted spin distributions. Here $k$ is the number of fitted
parameters. The uniform distribution has no fitted parameters because its support is fixed. Lower
AIC$_{\rm c}$ values are preferred.}
\label{tab:models}
\begin{tabular}{lrrr}
\toprule
Model & $k$ & $\log L$ & AIC$_{\rm c}$ \\
\midrule
Tapered uniform + Gaussian & 4 & -240.94 & 491.05 \\
Skew-normal & 3 & -243.77 & 494.22 \\
Two Gaussians & 5 & -242.36 & 496.55 \\
Fixed uniform & 0 & -249.48 & 498.96 \\
Gaussian & 2 & -250.28 & 504.89 \\
\bottomrule
\end{tabular}
\end{table}

\section{Mixture Parameters}\label{app:mixture_par}
\begin{table*}[ht!]
\centering
\small
\caption{Mixture parameters and central 95\% parametric-bootstrap intervals. Note that the endpoint of the tapered component is not the same quantity as the cutoff used in~\citet{2017ApJ...850..106P}.}
\label{tab:model_params}
\begin{tabular}{llrr}
\toprule
Model & Parameter & Estimate & 95\% interval \\
\midrule
Tapered uniform + Gaussian & Gaussian fraction & $0.412$ & $[0.176, 0.590]$ \\
Tapered uniform + Gaussian & Fast mean (Hz) & $565.7$ & $[542.7, 596.4]$ \\
Tapered uniform + Gaussian & Fast width (Hz) & $32.4$ & $[7.4, 50.3]$ \\
Tapered uniform + Gaussian & Taper endpoint (Hz) & $487.5$ & $[385.9, 579.6]$ \\
Two Gaussians & Slow fraction & $0.632$ & $[0.395, 0.810]$ \\
Two Gaussians & Slow mean (Hz) & $329.4$ & $[259.4, 380.9]$ \\
Two Gaussians & Slow width (Hz) & $114.7$ & $[64.2, 147.7]$ \\
Two Gaussians & Fast mean (Hz) & $568.3$ & $[535.1, 590.3]$ \\
Two Gaussians & Fast width (Hz) & $31.4$ & $[11.0, 54.8]$ \\
\bottomrule
\end{tabular}
\end{table*}

\section{Orbital Comparison}\label{app:orbit}
\begin{table*}[ht!]
\centering
\small
\begin{tabular}{lrrrrr}
\toprule
Sample & $\nu_{\rm cut}$ (Hz) & $n_{\rm F}$ & $n_{\rm S}$ & $\widehat A$ & $p$ \\
\midrule
Secure periods (current endpoint) & 487.5 & 10 & 19 & 0.847 & 0.0007821 \\
Secure periods (2017 boundary) & 540.0 & 7 & 22 & 0.750 & 0.02463 \\
Secure AMXPs only & 487.5 & 7 & 19 & 0.872 & 0.001307 \\
Including 4U\,1608--52 candidate & 487.5 & 11 & 19 & 0.861 & 0.0003126 \\
\bottomrule
\end{tabular}
\caption{Orbital-period comparison at the current spin endpoint $c$. All probabilities are nominal one-sided exact conditional permutation summaries for longer periods in the fast group, with observed ties retained.}
\label{tab:orbit_comparisons}
\end{table*}

\begin{table}[!ht]
\centering
\small
\caption{Sensitivity of the orbital-period comparison to
the adopted spin threshold.}
\label{tab:cutoff_sensitivity}
\begin{tabular}{rrrrc}
\toprule
$\nu_{\rm cut}$ (Hz) &
$n_{\rm F}$ &
$n_{\rm S}$ &
$\widehat{A}$ &
$p_{\mathrm{MW,orbit}}$ \\
\midrule
436.0 & 13 & 16 & 0.889 & $7.45\times10^{-5}$ \\
487.5 & 10 & 19 & 0.847 & $7.82\times10^{-4}$ \\
520.0 & 9 & 20 & 0.806 & $3.93\times10^{-3}$ \\
540.0 & 7 & 22 & 0.750 & $2.46\times10^{-2}$ \\
556.1 & 5 & 24 & 0.688 & $1.02\times10^{-1}$ \\
\bottomrule
\end{tabular}
\tablefoot{
Only the 29 systems with secure orbital-period estimates are included. Fast systems satisfy $\nu>\nu_{\rm cut}$ and slow systems satisfy $\nu\leq\nu_{\rm cut}$.
The primary division is 487.5 Hz. $\widehat{A}=U_{\rm fast}/(n_{\rm F}n_{\rm S})$, with ties contributing one half. All probabilities are one-sided exact permutation $p$-values for the alternative that fast systems tend to
have longer orbital periods.}
\end{table}

\section{Quiescent thermal constraints}
\label{app:thermal_catalogue}

\begin{table}[ht!]
\centering
\scriptsize
\renewcommand{\arraystretch}{1.12}
\caption{All 31 adopted quiescent thermal constraints.
\textbf{A} denotes a hydrogen-atmosphere estimate or bound, whereas \textbf{B} denotes a newly calculated blackbody temperature upper limit. MK means $10^6$~K. The symbol $<$ denotes a published upper limit and $\lesssim$ an approximate upper limit. 
$^a$~IGR~J16597 has a published 75-eV limit (0.67--0.87~MK). We use 0.87~MK conservatively.
$^b$~MAXI~J0911 has an upper limit derived from assigning all detected flux to a blackbody surface. Its flux uncertainty gives 0.76--0.89~MK.
$^c$~SAX~J1810 has a temperature conditional on an atmosphere plus power-law model, the thermal component is not required.
For HETE~J1900, the value is explicitly approximate.
In MAXI~J1957, the upper limit assumes a distance of 5~kpc and becomes 1.08--1.40~MK over the distance range 3--7~kpc. }
\label{tab:thermal_all}
\begin{tabular}{lrlcp{1.4cm}p{5.3cm}}
\hline\hline
Source & Spin (Hz) & $T^\infty$ (MK) & Method & Year & Reference \\
\hline
IGR J16597--3704 & 105 & $<0.87\,^{a}$ & A & 2008 & {\raggedright\citet{thermalTetarenko2018}\par} \\
Swift J1756.9--2508 & 182 & $\lesssim0.96$ & A & 2009 & {\raggedright\citet{gusakov2014,thermalHaskell2012}\par} \\
XTE J0929--314 & 185 & $<0.58$ & A & 2004 & {\raggedright\citet{thermalWijnands2005}\par} \\
XTE J1807--294 & 190 & $\lesssim0.45$ & A & 2004 & {\raggedright\citet{gusakov2014,thermalHeinke2009}\par} \\
NGC 6440 X--2 & 206 & $\lesssim0.37$ & A & 2000--2003 & {\raggedright\citet{gusakov2014,thermalHaskell2012}\par} \\
IGR J17511--3057 & 245 & $\lesssim1.10$ & A & not reported & {\raggedright\citet{gusakov2014,thermalHaskell2012}\par} \\
IGR J18245--2452 & 254 & $<0.51$ & A & 2002 & {\raggedright\citet{thermalLinares2014}\par} \\
IGR J17191--2821 & 294 & $\lesssim0.86$ & A & 2007 & {\raggedright\citet{gusakov2014,thermalHaskell2012}\par} \\
XTE J1814--338 & 314 & $\lesssim0.61$ & A & 2005 & {\raggedright\citet{gusakov2014,thermalHeinke2009}\par} \\
MAXI J1957+032 & 314 & $\lesssim1.25$ & B & 2015--2016 & {\raggedright\citet{thermalMataSanchez2017,thermalRavi2017}\par} \\
MAXI J0911--655 & 340 & $\lesssim0.84\,^{b}$ & B & 2007 & {\raggedright\citet{thermalServillat2008,thermalHoman2016}\par} \\
IGR J17494--3030 & 376 & $\lesssim1.15$ & B & 2012 & {\raggedright\citet{ArmasPadilla2013_17494,Chakrabarty2013_4886}\par} \\
HETE J1900.1--2455 & 377 & $\simeq 0.41$ & A & 2018 & {\raggedright\citet{thermalDegenaar2021}\par} \\
IGR J17498--2921 & 401 & $\lesssim0.93$ & A & 2006 & {\raggedright\citet{gusakov2014,thermalDegenaar2012}\par} \\
SAX J1808.4--3658 & 401 & $<0.27$ & A & 2001--2007 & {\raggedright\citet{thermalHeinke2009}\par} \\
XTE J1751--305 & 435 & $\lesssim0.63$ & A & 2004 & {\raggedright\citet{gusakov2014,thermalHeinke2009}\par} \\
SAX J1748.9--2021 & 442 & $1.06^{+0.02}_{-0.02}$ & A & 2009 & {\raggedright\citet{thermalBahramian2015}\par} \\
SRGA J144459.2--604207 & 448 & $\lesssim2.12$ & B & 2010 & {\raggedright\citet{SidoliSguera2024_16470,Ng2024_SRGA_thermal}\par} \\
Swift J1749.4--2807 & 518 & $<1.16$ & A & 2011 & {\raggedright\citet{thermalDegenaar2012}\par} \\
KS 1731--260 & 524 & $0.75\pm0.01$ & A & 2015 & {\raggedright\citet{thermalMerritt2016}\par} \\
IGR J17591--2342 & 527 & $\lesssim2.53$ & B & 2018 & {\raggedright\citet{Gusinskaia2020_17591}\par} \\
MAXI J1816--195 & 528 & $\lesssim2.49$ & B & 2022 & {\raggedright\citet{thermalMarino2022}\par} \\
SAX J1810.8--2609 & 532 & $0.64\pm0.04\,^{c}$ & A & 2003 & {\raggedright\citet{thermalJonker2004}\par} \\
Aql X--1 & 550 & $1.16\pm0.02$ & A & 2012 & {\raggedright\citet{thermalWaterhouse2016}\par} \\
EXO0748--676 & 552 & $1.326\pm0.001$ & A & 2018 & {\raggedright\citet{thermalParikh2020}\par} \\
MXB 1659--298 & 567 & $0.65\pm0.05$ & A & 2018 & {\raggedright\citet{thermalParikh2019}\par} \\
4U 1730--22 & 585 & $1.16\pm0.18$ & A & 2004 & {\raggedright\citet{thermalTomsick2007}\par} \\
GRS 1741.9--2853 & 589 & $\lesssim2.74$ & B & 1999 & {\raggedright\citet{thermalMuno2003}\par} \\
IGR J00291+5934 & 599 & $0.63^{+0.04}_{-0.08}$ & A & 2005 & {\raggedright\citet{thermalHeinke2009}\par} \\
SAX J1750.8--2900 & 601 & $<1.23$ & A & 2016 & {\raggedright\citet{thermalParikh2017}\par} \\
4U 1608--52 & 619 & $1.51\pm0.27$ & A & 1993 & {\raggedright\citet{thermalRutledge1999}\par} \\
\hline
\end{tabular}
\end{table}

\end{appendix}

\end{document}